\documentclass[superscriptaddress,prl,english,floatfix,twocolumn,10pt]{revtex4-2}
\usepackage{graphicx}
\usepackage{float}
\usepackage{natbib}
\usepackage{amssymb}
\usepackage{amsmath}
\usepackage{bbold}
\usepackage{mathtools}
\usepackage[utf8]{inputenc}
\usepackage{braket}
\usepackage{graphicx}
\usepackage{subfigure}
\usepackage{pgfplots}
\usepackage{csquotes}
\usepackage{hhline}
\usepackage{amssymb}

\usepackage{dcolumn}
\usepackage{tabularx}
\usepackage[colorlinks=true,linkcolor=blue,citecolor=blue,urlcolor=blue]{hyperref}
\usepackage{longtable}
\usepackage{braket}
\usepackage{float}
\usepackage{listings}
\usepackage{color}
\definecolor{mygreen}{rgb}{0,0.6,0}
\definecolor{mygray}{rgb}{0.5,0.5,0.5}
\definecolor{mymauve}{rgb}{0.58,0,0.82}
\graphicspath{ {./IMAGE1/} }
\newcolumntype{C}{>{\centering\arraybackslash}X}

\begin{document}

\preprint{APS/123-QED}

\title{Engineering unique localization transition with coupled Hatano-Nelson chains}

\author{Ritaban Samanta}
\email{ritabansamanta28@gmail.com}
\author{Aditi Chakrabarty}%
\email{aditichakrabarty030@gmail.com}
\author{Sanjoy Datta}
\email{dattas@nitrkl.ac.in}
\affiliation{%
 Department of Physics and Astronomy, National Institute of Technology, Rourkela, Odisha-769008, India.
}%

\date{\today}

\begin{abstract}
The Hatano-Nelson (HN) Hamiltonian has played a pivotal role in catalyzing research
interest in non-Hermitian systems, primarily because it showcases unique physical phenomena that arise solely due to non-Hermiticity. The non-Hermiticity in the HN Hamiltonian, driven by asymmetric hopping amplitudes, induces a delocalization-localization (DL) transition in a one-dimensional (1D) lattice with random disorder,
sharply contrasting with its Hermitian counterpart. A similar DL transition occurs in a
1D quasiperiodic HN (QHN) lattice, where a critical quasiperiodic potential strength 
separates metallic and insulating states, akin to the Hermitian case. In these systems,
all states below the critical potential are delocalized, while those above are localized. In this study, we reveal that coupling two 1D QHN lattices can significantly
alter the nature of the DL transition. We identify two critical points, 
$V_{c1} < V_{c2}$, when the nearest neighbors of the two 1D QHN lattices are 
cross-coupled with strong hopping amplitudes under periodic boundary conditions (PBC).
Generally, all states are completely delocalized below $ V_{c1}$ and completely localized above $V_{c2}$, while two mobility edges symmetrically emerge about $\rm{Re[E]} = 0$ between $V_{c1}$ and $V_{c2}$. Notably, under specific asymmetric
cross-hopping amplitudes, $V_{c1}$ approaches zero, resulting in localized states even for \emph{infinitesimally weak potential}. Remarkably, we also find that the mobility edges precisely divide the delocalized and localized states in \emph{equal proportions.} Furthermore, we observe that the conventional one-to-one correspondence between electronic states under PBC and open boundary conditions (OBC) in 1D HN lattices breaks down within certain regions of the parameter space of the coupled QHN system.

\end{abstract}

\maketitle

\section{Introduction}\label{sec:ntroduction}  

The concept of localization of the matter waves was laid down by P.W. Anderson in 1958, wherein the investigation revealed that in the presence of a sufficiently strong random disorder in the 3D lattice, the electronic conductivity ceases, hence becoming an insulator (frequently termed as the $Anderson$ localization) \citep{Anderson}.
The interesting features of Anderson localization has been implemented in many domains of physics, such as superconductors \citep{katsanos1998superconductivity,burmistrov2012enhancement,leavens1985anderson}, photonics \citep{sarychev1999anderson,lahini2008anderson,jovic2011anderson,qiao2019cavity} and acoustics \citep{condat1987observability,cohen1987crossover}.
However, it was later demonstrated using a scaling law that in the lattices of lower dimensions (1D/2D), even an infinitesimally small strength of the random disorder localizes all the electronic wave functions \citep{abrahams1979scaling}.
A few years later, in 1980, S. Aubry and G. Andr\'e demonstrated that in quasiperiodic lattices, a delocalization-localization (DL) transition takes place even in lower dimensions \cite{aubry1980analyticity,harper1955single}.
In the cosine-modulated Aubry-André-Harper (AAH) models, the DL transition occurs at a finite value of the quasiperiodic potential, governed by the self-duality of the Hamiltonian in the real and momentum spaces \cite{longhi2019metal}.
For closed quantum systems which are described by Hermitian Hamiltonians, there have been many works based on the AAH model in the last few years \citep{tong1994localization,ossipov2001quantum,yoo2020nonequilibrium,Sokoloff,sokoloff1985unusual,kohmoto1983metal,Thouless,hiramoto1992electronic,kramer1993localization,chang1997multifractal}.
Recently, such quasiperiodic lattices have been realized in the ultracold atomic systems \cite{roati2008anderson,deissler2010delocalization,schreiber2015observation}.\\
\indent However, in reality, most of the condensed matter systems are coupled to the environment that exchanges either energy, or particles, or both with the surroundings.
Such open systems are frequently mapped using a non-Hermitian Hamiltonian.
Hatano and Nelson in 1996 introduced one such model which is an extension of the Anderson model with asymmetric hopping amplitudes.
In his work, originally on the superconductors, it was shown that in the presence of such random disorder, the DL transition is manifested in 1D systems.
There have been many ongoing studies on the localization, spectral properties, self-duality and mobility edges in various non-Hermitian systems \citep{tang2021localization,tzortzakakis2021transport,zeng2017anderson,zhai2020many,yoo2020nonequilibrium}.
Besides, such systems with asymmetric hopping amplitudes have been gaining attention over the years due to the phenomenon of skin effect wherein a macroscopic number of bulk states become localized at one of the edges under open boundaries \cite{Lee,Torres,Lee_2019}.
\\
\indent On the other hand, some recent works have been carried out on coupled AAH chains in which two disparate chains of atoms are coupled to each other by some interchain hopping amplitudes \citep{PhysRevB.105.205402,PhysRevB.99.054211}.
It was demonstrated that such a coupled Hermitian AAH chain shows interesting properties like the existence of mobility edges.
However, to the best of our knowledge, the interplay of the quasiperiodicity and the coupling between the non-Hermitian chains of Hatano-Nelson(HN) type have not been investigated so far.
Therefore, the aim of this work is to investigate a coupled HN bipartite chain in the presence of AAH type potential to closely scrutinize the localization behavior in such coupled systems.
Intriguingly, we find that the presence of a strong interchain coupling between two dissimilar atoms in the two sublattices possessing symmetric and asymmetric interchain hopping between two atoms of adjacent unit cells render equal proportion of localized and delocalized states in the presence of quasiperiodic potential.
Moreover, we find that in the latter case, half of the states are localized even for a very weak strength of the quasiperiodic potential, akin to the study by Anderson on 1D systems.
Finally, we reveal that the coupling renders distinct properties in the skin effect as compared to the conventional HN systems, wherein some of the localized states (in the bulk) under PBC become skin-states under the OBC.\\
\indent This work is organized as follows: In Sec.\ref{sec:hamiltonian}, we discuss the coupled QHN Hamiltonian and elaborate the method to numerically identify the delocalized and localized phases in Sec.~\ref{sec:IPR}.
We analytically determine the strength of the quasiperiodic potential ($V_{c,1}$ and $V_{c,2}$) where the localization transitions occur in Sec.~\ref{sec:ANALYTICAL}.
In Sec.~\ref{sec:Results}, we demonstrate our unique findings in the presence of various ratios of the strong interchain coupling between the two QHN chains.
We propose a feasible experimental set-up in coupled optical waveguides in Sec.~\ref{sec:Expt}.
Finally, Sec.~\ref{sec:conclusion} consists of a summary of the work, highlighting the important results and unique findings.

\section{\label{sec:model}Model and methods}
\subsection{\label{sec:hamiltonian}The coupled QHN Hamiltonian}

We consider two uni-directional HN chains with quasiperiodic potential (consisting of two sublattices 
A and B in a single unit cell) coupled to each other $via.$ an interchain
hopping, which we call a coupled quasiperiodic HN (QHN) Hamiltonian from here on.
 \begin{figure}[]
 	\centering
 	\includegraphics[width=0.4 \textwidth,height=0.24\textwidth]{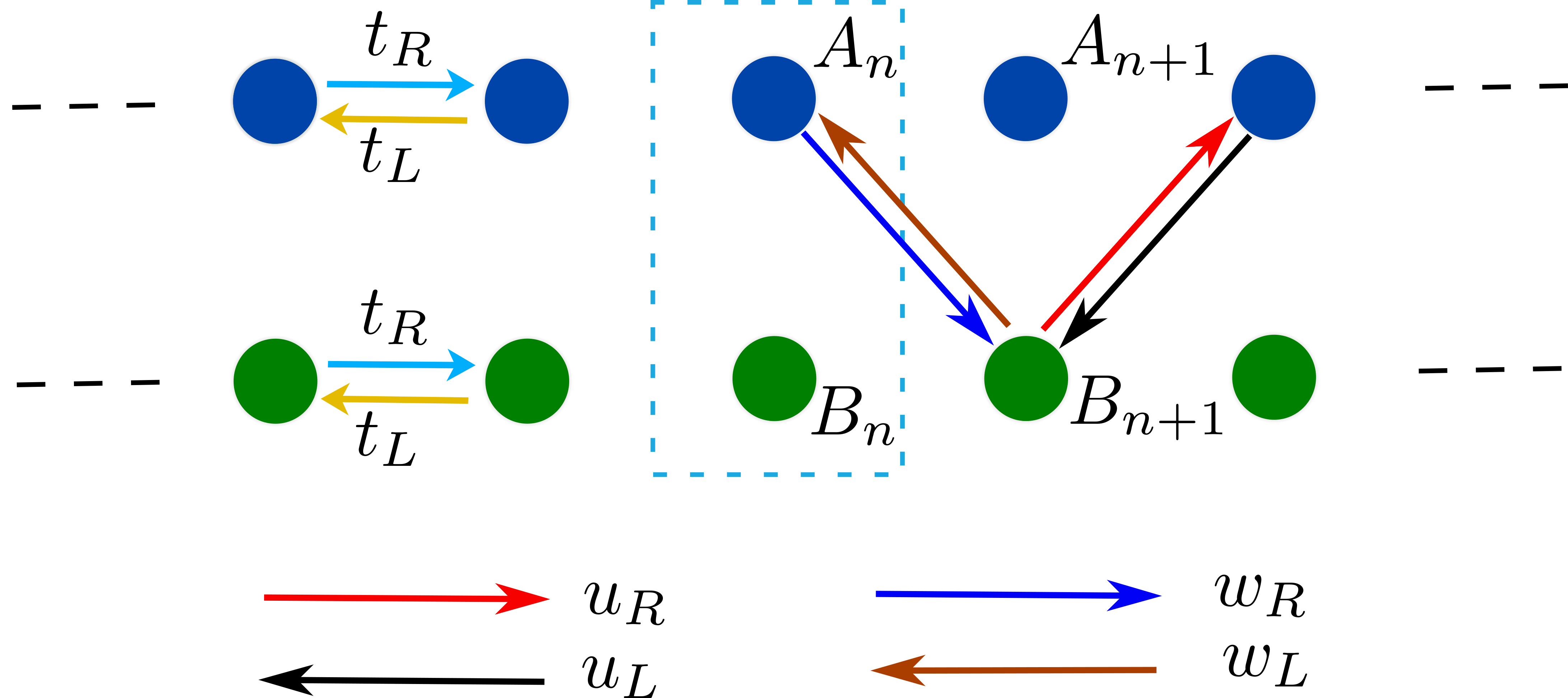}	
 	\caption{Schematic diagram of the coupled QHN model. Atoms $A$ are depicted in blue and atoms $B$ are depicted in green.The n$th$ unit cell containing the two atoms is demonstrated by the dashed rectangle. The different interchain hopping amplitudes are mentioned below and are represented by coloured arrow lines.}
 	\label{Fig:1}
 \end{figure}

 	The Hamiltonian in such a coupled system is given by,
 	\begin{equation}
 		\label{eqn:1}
 		{\mathcal{H}}={\mathcal{H}}_{A}+{\mathcal{H}}_{B}+{\mathcal{H}}_{C},
 	\end{equation}
 	where, ${\mathcal{H}}_{A}({\mathcal{H}}_{B})$ is the Hamiltonian for chain 1(2) 
 	of atom A(B) and ${\mathcal{H}}_{C}$ introduces the interchain coupling between chains 1 and 2. 
 	The individual terms of the Hamiltonian are described as,
 	\begin{eqnarray}
 		{\mathcal{H}}_{A(B)}=\sum_{n=1}^{N-1}\Big(t_{R}{{c}}_{n+1,A(B)}^\dag c_{n,A(B)}+ t_{L}{{c}}_{n,A(B)}^\dag c_{n+1,A(B)}\Big)\nonumber \\
 		+\sum_{n=1}^{N}V \text{cos}(2\pi n \alpha){{c}}_{n,A(B)}^\dag c_{n,A(B)}.~~~~~~~~~~~~~~\label{eqn:2}
 	\end{eqnarray}
 
	Here, ${{c}}_{n,x}^\dag({c}_{n,x})$  are the fermionic creation (annihilation) operators at the site $n$ of sublattice $x=A(B)$.
	The first two terms of
	the Hamiltonian $H_{A(B)}$ define the usual asymmetric intrachain hopping of the fermions
	between the nearest neighbour sites in sublattices $A(B)$ and 
	the second term is the onsite quasiperiodic potential.
	$\alpha$ is an irrational number approximated as $F_{n-1}/F_{n}$, where $F_{n}$ and $F_{n-1}$ are the n$th$ and (n-1)$th$ terms of the Fibonacci series respectively.
	Throughout this work, we have considered $\alpha$ to be $(\sqrt{5}-1)/2$ which approximates the inverse golden mean ratio.
	The final part of the Hamiltonian which couples the two distinct HN chains $via.$ interchain coupling amplitudes is given as,
	\begin{eqnarray}
	{\mathcal{H}}_{C} =\sum_{n=1}^{N}\Big(u_{R}{{c}}_{n+1,A}^\dag c_{n,B}+u_{L}{{c}}_{n,B}^\dag c_{n+1,A}\nonumber \\
	+w_{R}{{c}}_{n+1,B}^\dag c_{n,A}+w_{L}{{c}}_{n,A}^\dag c_{n+1,B}\Big). \label{eqn:4}
	\end{eqnarray}

	The interchain coupling $u_{R}(u_{L})$ is the hopping strength from $B_{n}\rightarrow A_{n+1}(A_{n+1}\rightarrow B_{n})$, whereas
	$w_{R}(w_{L})$ is the hopping strength of $A_{n}\rightarrow B_{n+1}(B_{n+1}\rightarrow A_{n})$.
	All these terms of the inter and intra chain coupling are depicted in a schematic in Fig.~\ref{Fig:1}.

	\subsection{\label{sec:IPR}Delocalization-localization (DL) transition: the IPR}

	\par The localized and delocalized behaviour of the eigenstates of the system is characterised by estimating the value of the Inverse Participation Ratio (IPR).
	The IPR for a given eigenstate ($m$) is given by \citep{PhysRevB.105.214203},

	\begin{equation}
	IPR_{m}=\frac{\sum_{n=1}^{N}\sum_{x=A,B}^{}|{\psi^{m}}_{n,x}|^{4}}{(\sum_{n=1}^{N}\sum_{x=A,B}^{}|{\psi^{m}}_{n,x}|^{2})^{2}}
	\label{eqn:5}
	\end{equation} 
	where, ${\psi^{m}}_{n,x}$ is the normalized wave function of eigenstate labelled by $m$ at site $n$ for the chain $x=A,B$. Here, $N$ is the size of the system and the number of total eigenstates is given by $L=2N$.
	It is well known that for the delocalized states, the $IPR$ varies as $IPR \sim L^{-1}$. In the thermodynamic limit ($N \rightarrow \infty$), and therefore $IPR \sim0$. In contrast, for the localized states, the $IPR$ is independent of the system size and approaches 1 in the thermodynamic limit.
	All our numerical estimates are for a lattice with 610 sites, unless specifically mentioned.
	\begin{figure*}[]
		\centering
		\includegraphics[width=0.23\textwidth,height=0.21\textwidth]{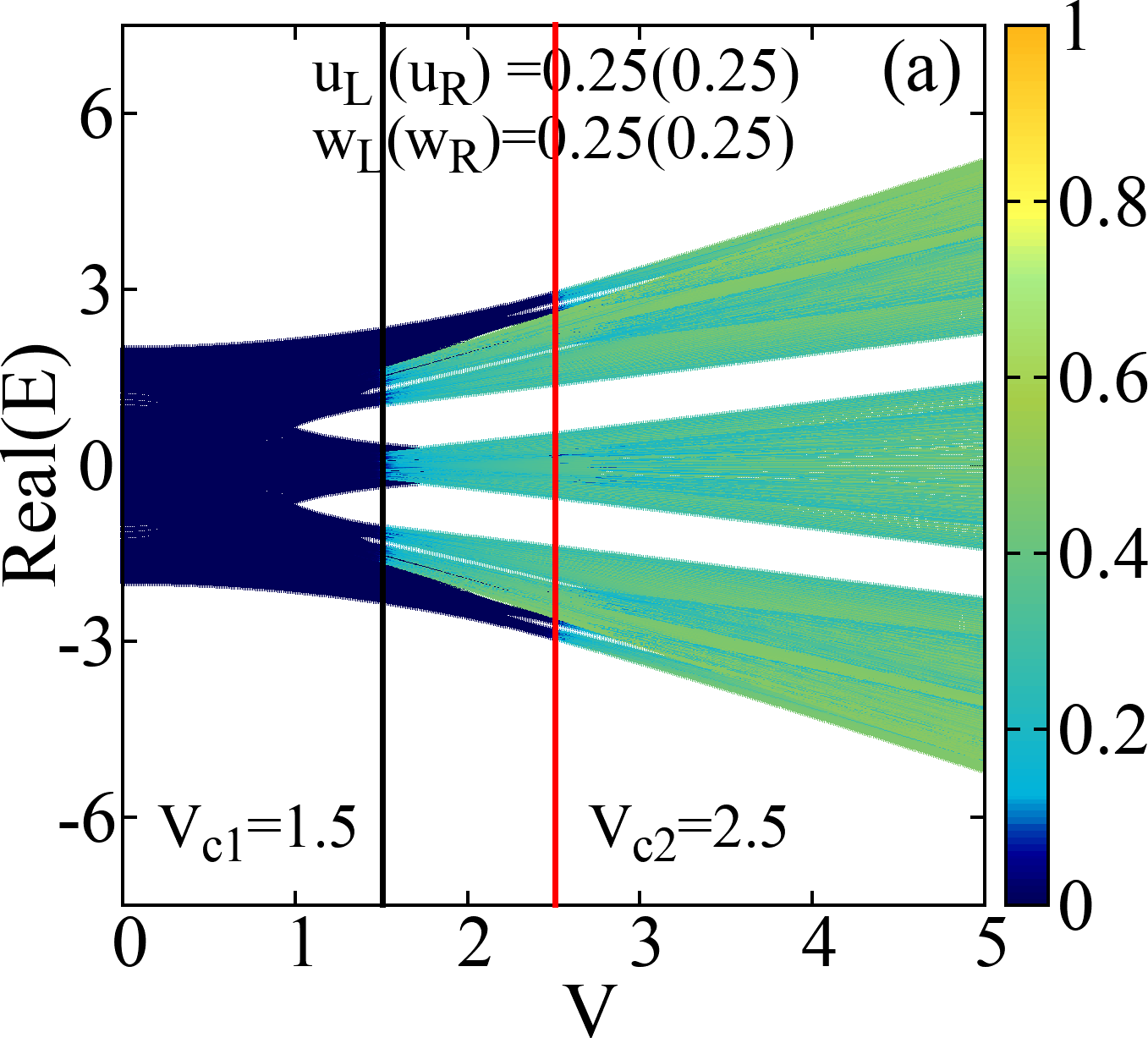}\hspace{0.1cm}
		\includegraphics[width=0.23\textwidth,height=0.21\textwidth]{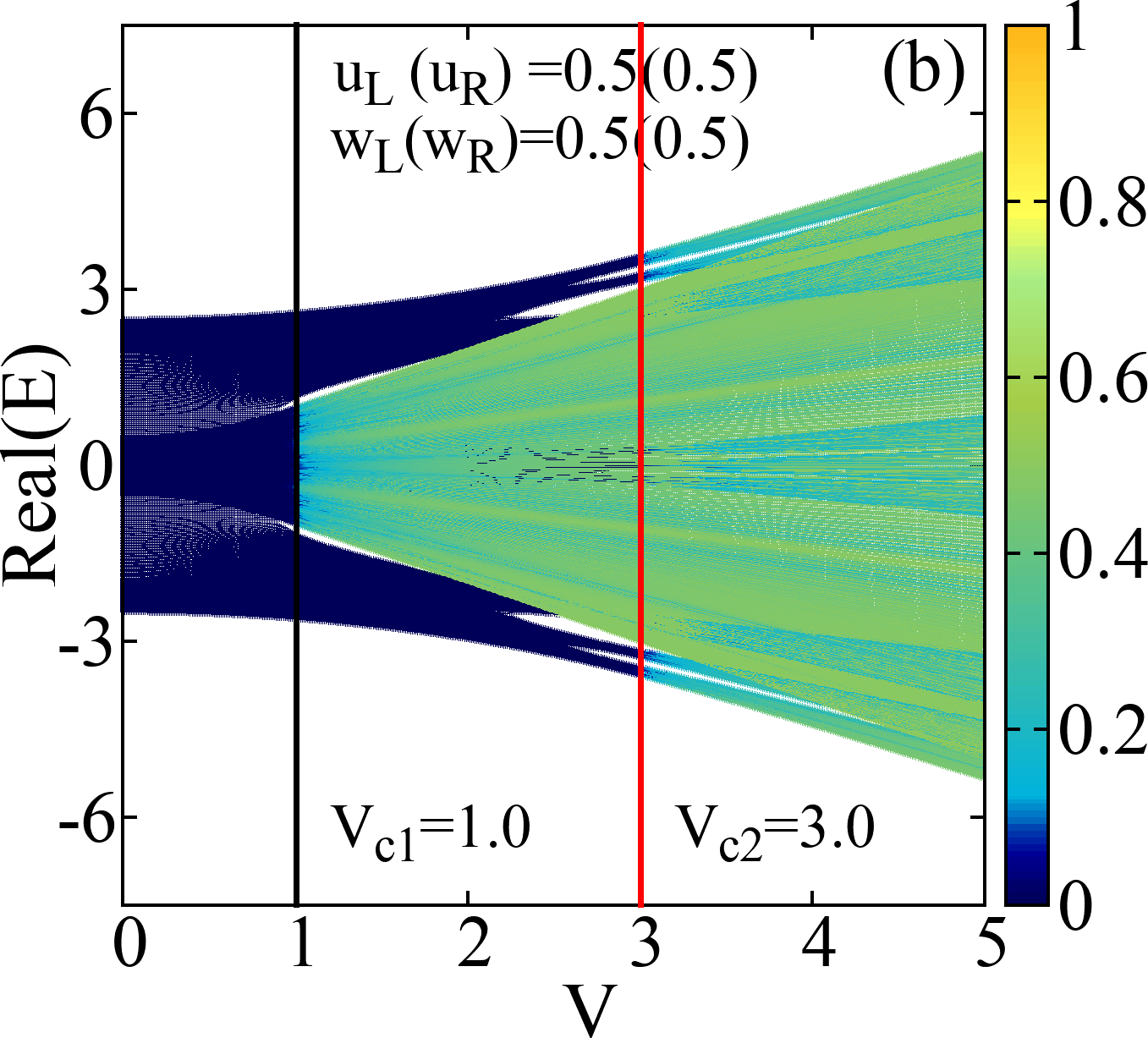}\hspace{0.1cm}
		\includegraphics[width=0.23\textwidth,height=0.21\textwidth]{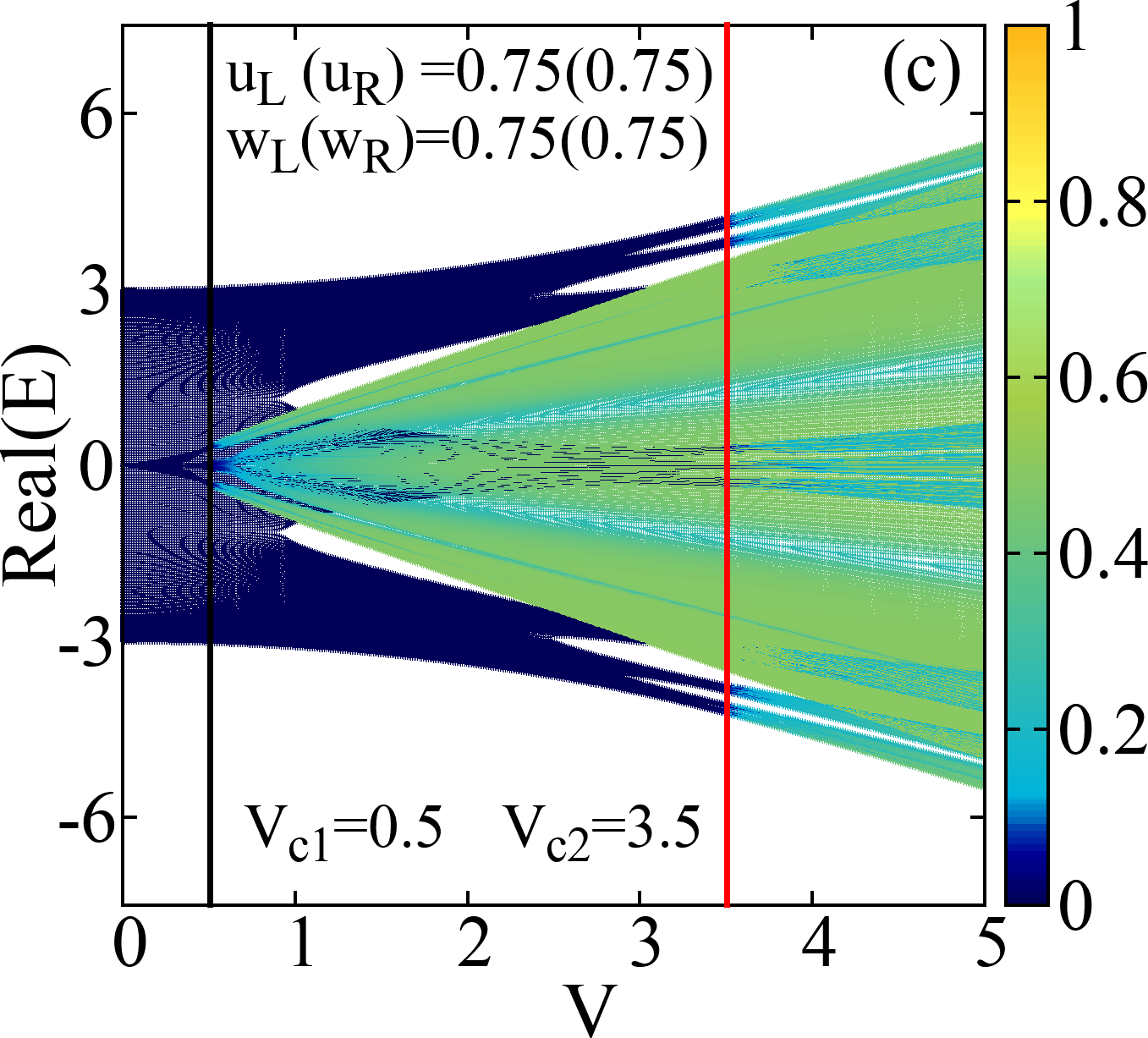}\hspace{0.1cm}
		\includegraphics[width=0.23\textwidth,height=0.21\textwidth]{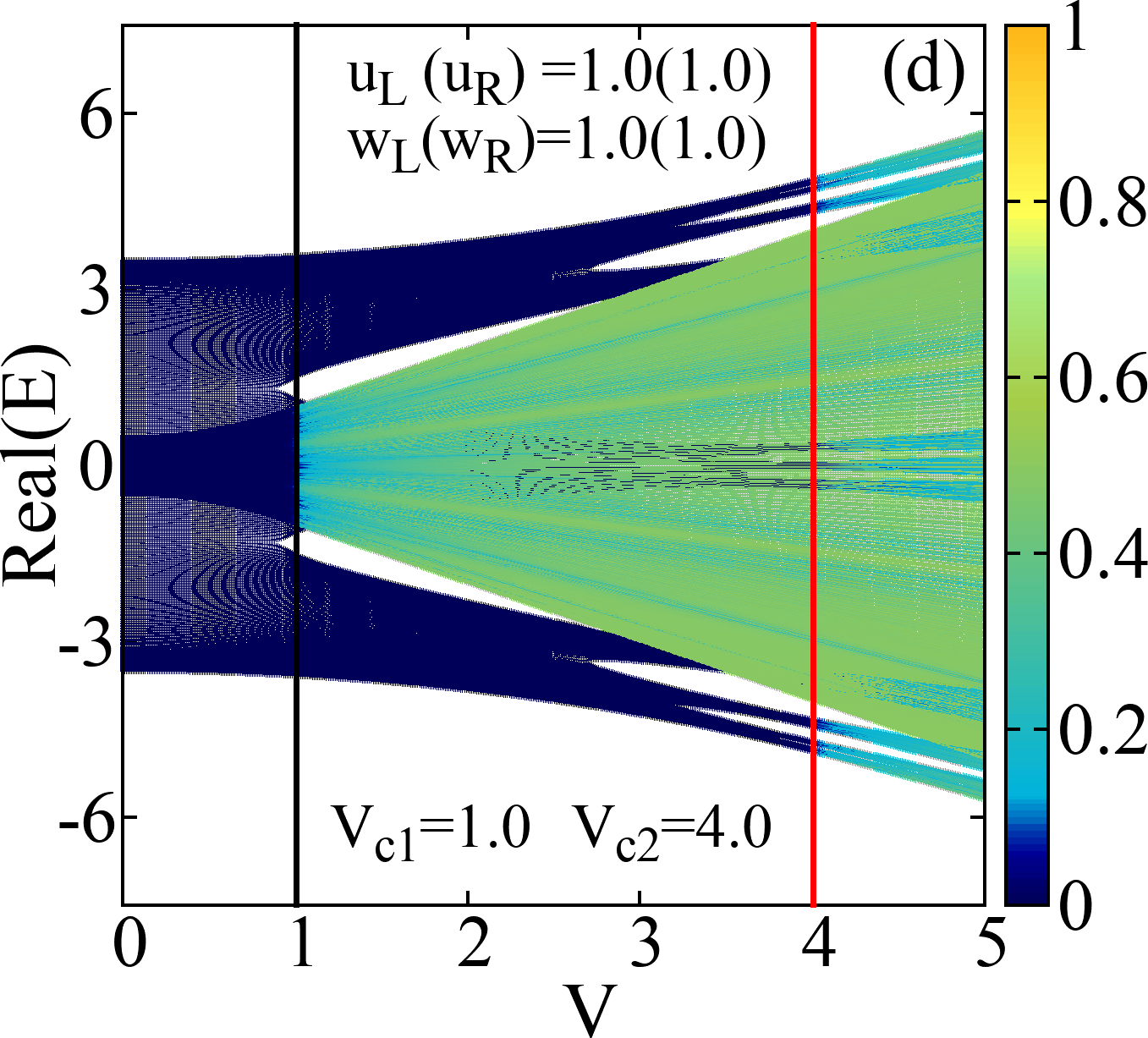}\hspace{0.1cm}
		\includegraphics[width=0.23\textwidth,height=0.21\textwidth]{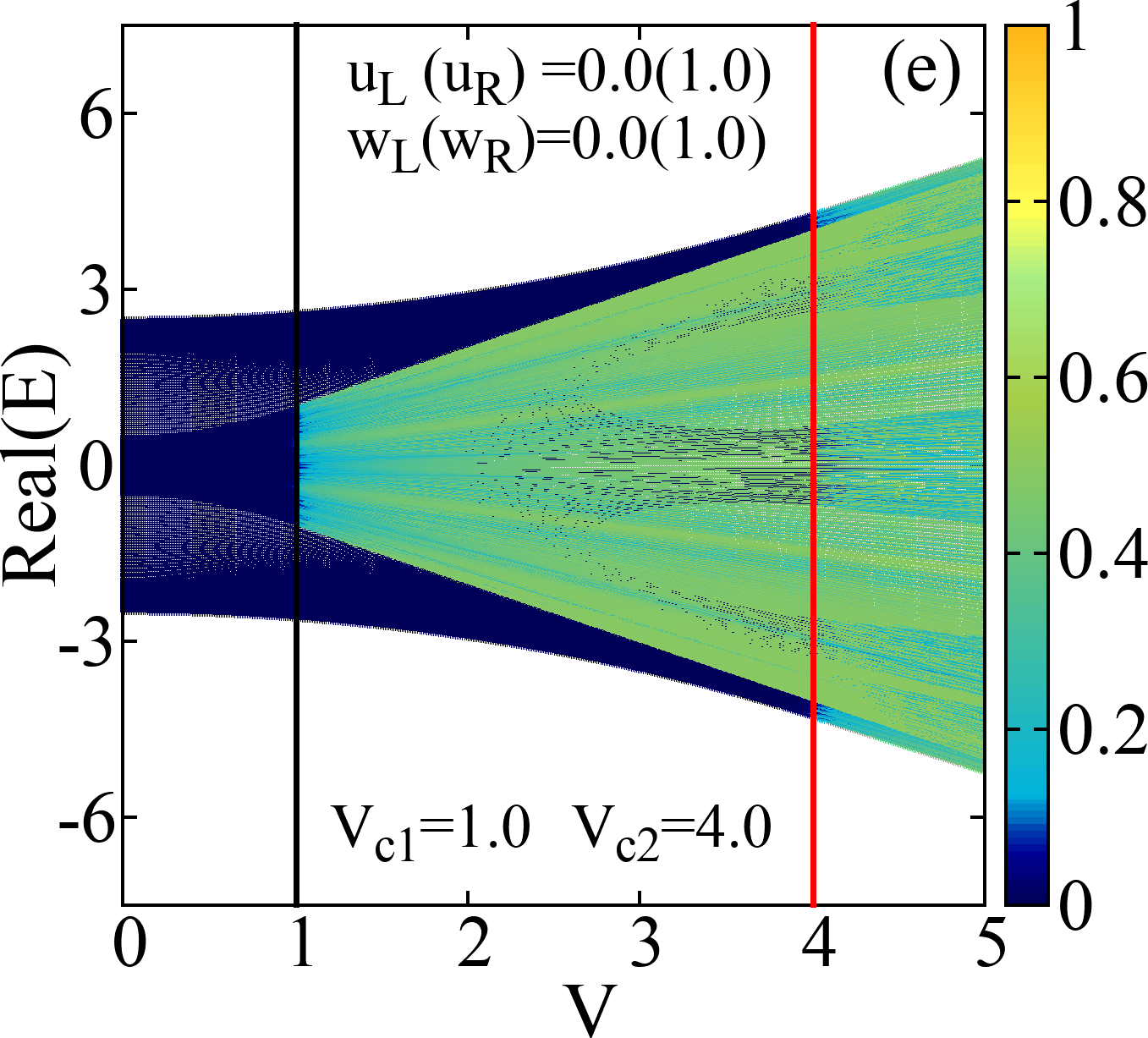}\hspace{0.1cm}
		\includegraphics[width=0.23\textwidth,height=0.21\textwidth]{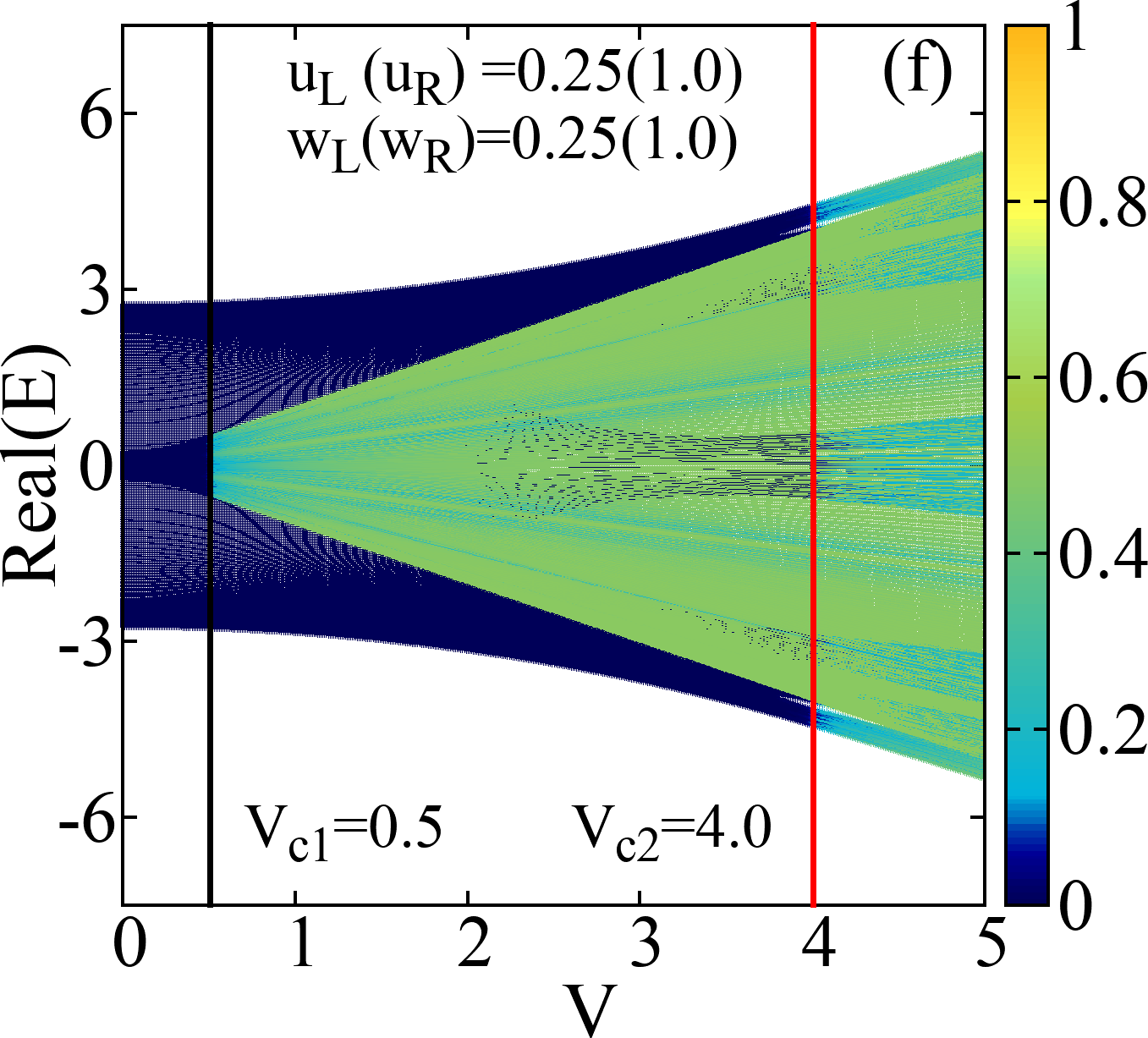}\hspace{0.1cm}
		\includegraphics[width=0.23\textwidth,height=0.21\textwidth]{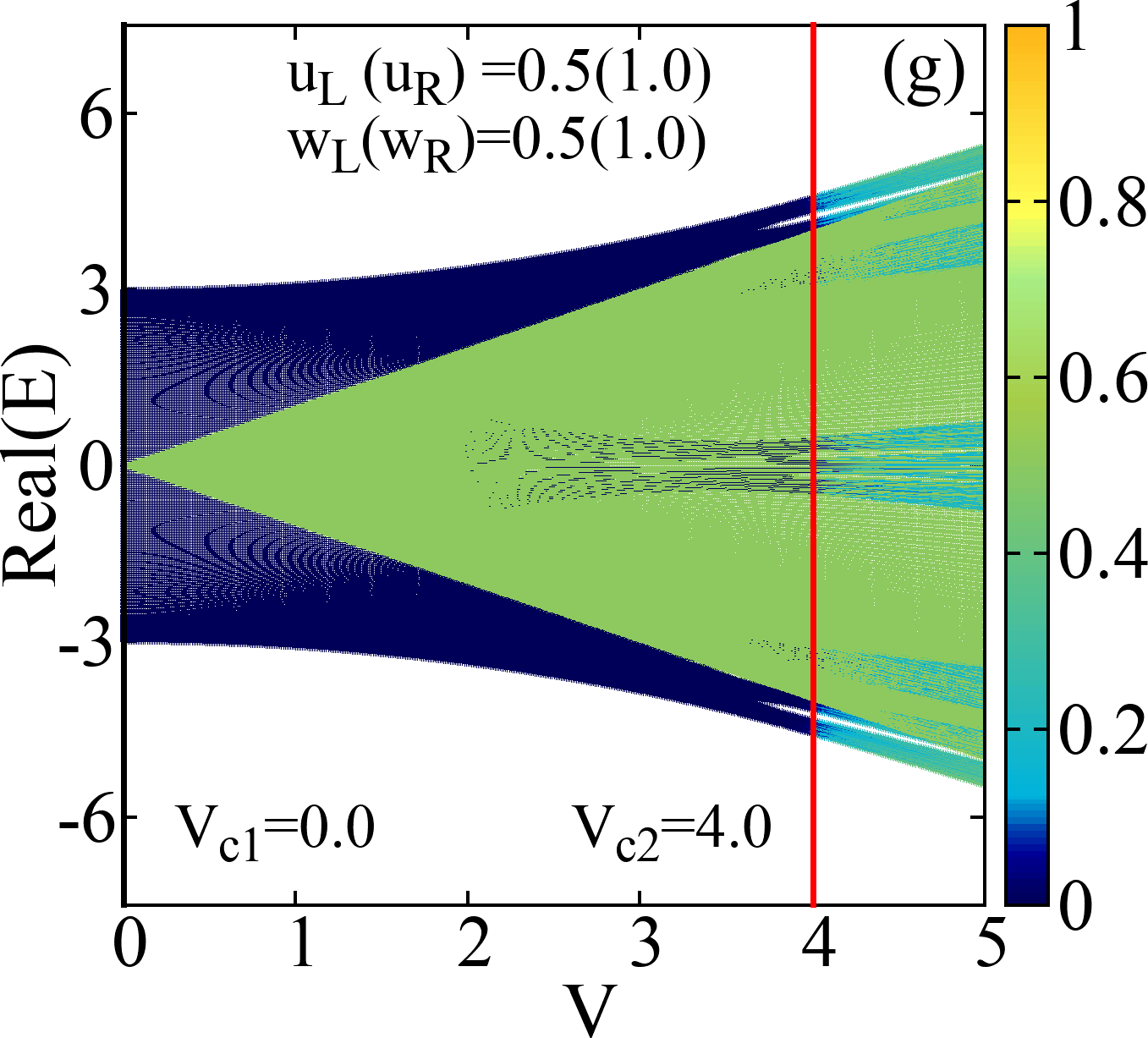}\hspace{0.1cm}
		\includegraphics[width=0.23\textwidth,height=0.21\textwidth]{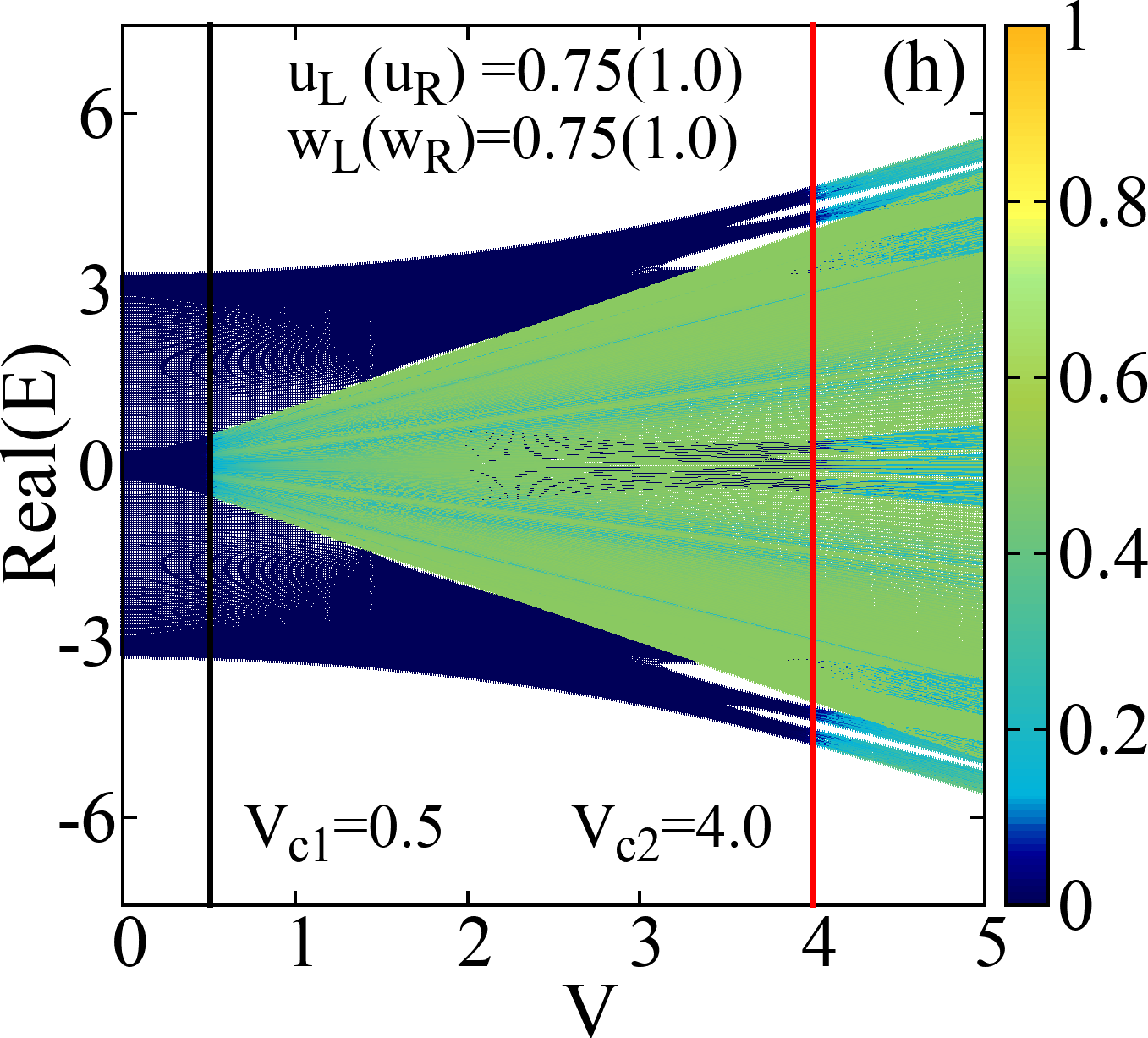}\hspace{0.1cm}
		\caption{The localization behavior in a strongly coupled QHN Hamiltonian with $t_L=0.5,t_R=1.0$ in a lattice with $N=610$ sites under PBC with interchain hopping between the two adjacent unit cells. Projection of IPR as a function of the real part of the eigen-energy and quasiperiodic potential ($V$) for (a-d) symmetric interchain hopping and (e-h) asymmetric interchain hopping. In both the panels, the DL transition is shown as a dark blue to green transition. In particular, the parameters of interchain coupling are: (a) $u_{L}=u_{R}=w_{L}=w_{R}=0.25$, (b) $u_{L}=u_{R}=w_{L}=w_{R}=0.5$, (c) $u_{L}=u_{R}=w_{L}=w_{R}=0.75$, (d) $u_{L}=u_{R}=w_{L}=w_{R}=1.0$, (e)$u_{L}=w_{L}=0.0~\text{and}~u_{R}=w_{R}=1.0$, (f)$u_{L}=w_{L}=0.25~\text{and}~u_{R}=w_{R}=1.0$, (g)$u_{L}=w_{L}=0.5~\text{and}~u_{R}=w_{R}=1.0$ (h)$u_{L}=w_{L}=0.75~\text{and}~u_{R}=w_{R}=1.0$.}
		\label{Fig:2}
	\end{figure*}

   \section{\label{sec:ANALYTICAL}Analytical understanding of the localization transition}
    In the following discussion, we analytically estimate the critical value of quasiperiodic potential for the DL transition in the coupled QHN Hamiltonian as defined in Sec. \ref{sec:hamiltonian}. The Hamiltonian consists of creation and annihilation operators of two sublattices, which can be effectively combined into a single equation in terms of a spinor representation \cite{Rossignolo} given as,
    \begin{gather}
      b=  \begin{pmatrix}
         c_A\\
          c_B
        \end{pmatrix}
    \label{Eq:spinor}
    \end{gather}
    Using Eq.~(\ref{Eq:spinor}), one can immediately obtain,
    \begin{eqnarray}
	{\mathcal{H}} =\sum_{n=1}^{N}\Big({{b}}_{n}^\dag T_1 b_{n+1}+{{b}}_{n+1}^\dag T_2 b_{n}\Big)+\sum_{n=1}^{N}{{b}}_{n}^\dag \epsilon(n) b_{n}, \label{eqn:9}
	\end{eqnarray}
 where,
 \begin{gather}
      \epsilon(n)=  \begin{pmatrix}
         V_n & 0\\
          0 & V_n
        \end{pmatrix}
    \end{gather}
    and
    \begin{gather}
      T_1=  \begin{pmatrix}
         t_{L} & w_L\\
          u_L & t_{L}
        \end{pmatrix}; T_2=  \begin{pmatrix}
         t_{R} & u_R\\
          w_R & t_{R}
        \end{pmatrix}.
    \end{gather}
			
	\begin{figure*}[]
		\centering
		\includegraphics[width=0.23 \textwidth,height=0.19\textwidth]{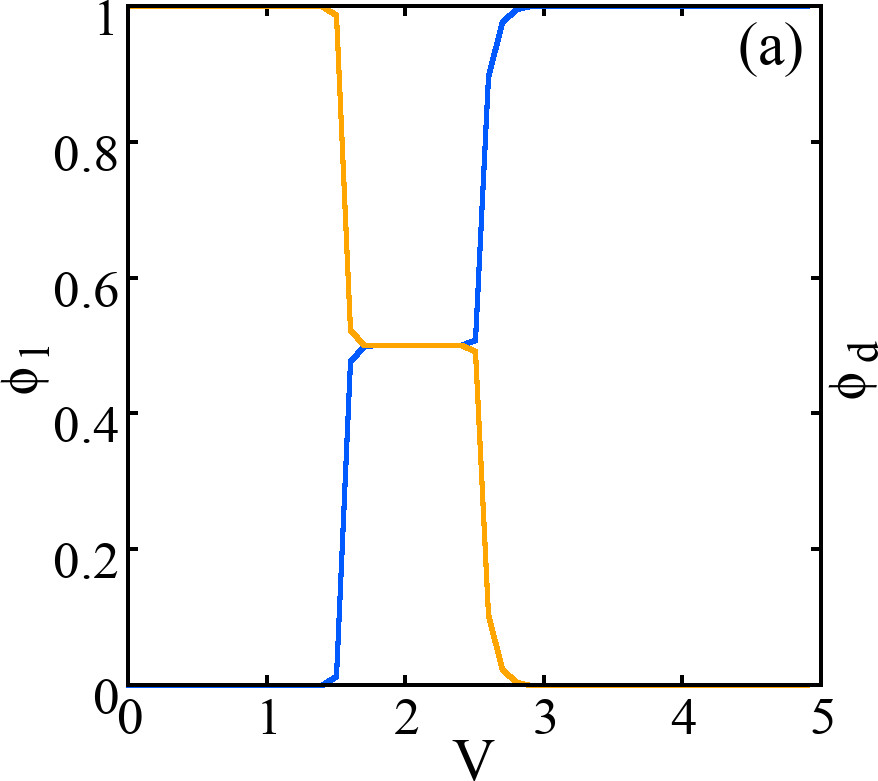}\hspace{0.2cm}
		\includegraphics[width=0.23 \textwidth,height=0.19\textwidth]{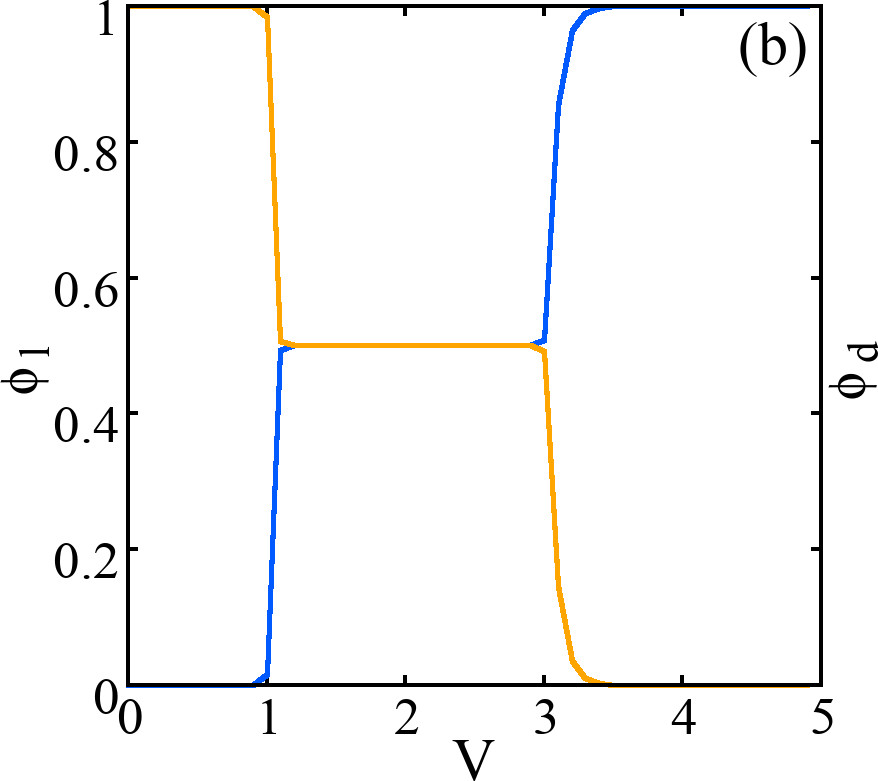}\hspace{0.2cm}
		\includegraphics[width=0.23 \textwidth,height=0.19\textwidth]{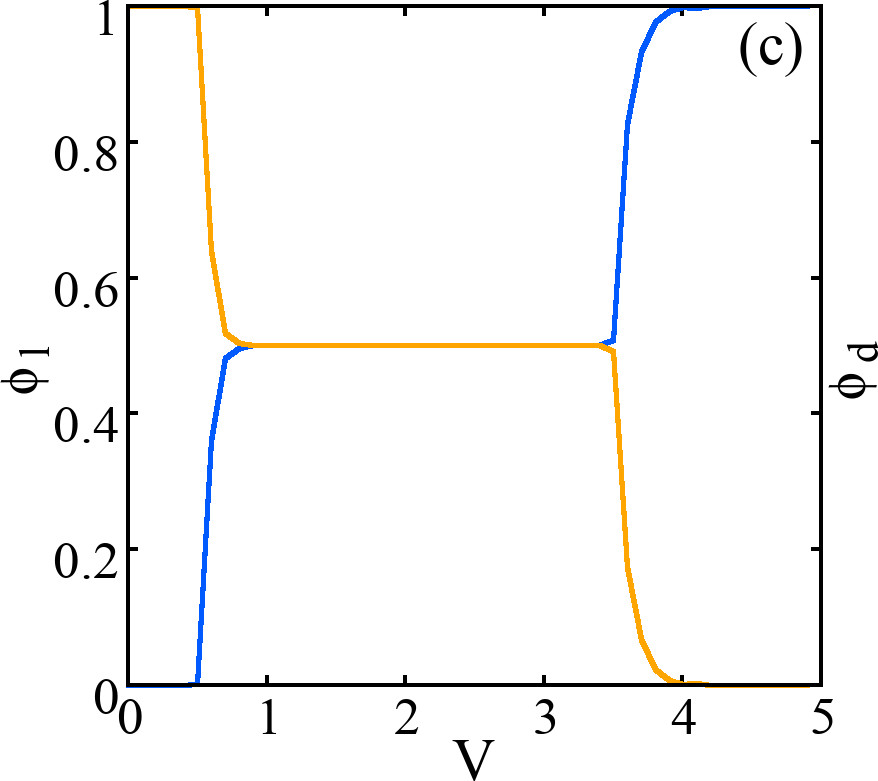}\hspace{0.2cm}
		\includegraphics[width=0.23 \textwidth,height=0.19\textwidth]{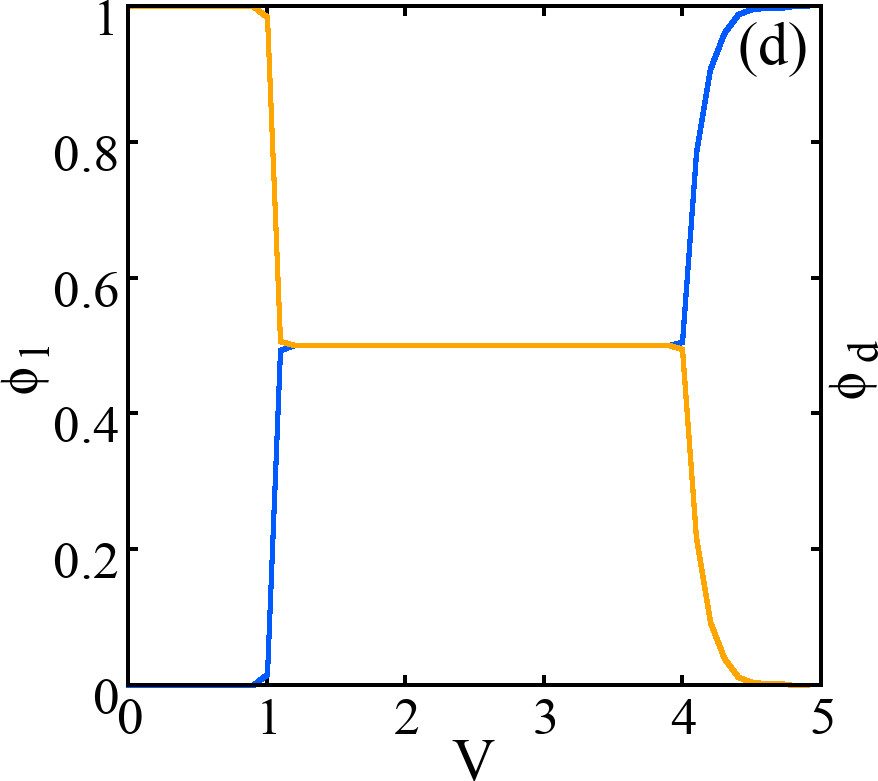}\hspace{0.2cm}\\
		\vspace{0.4cm}
		\includegraphics[width=0.23 \textwidth,height=0.19\textwidth]{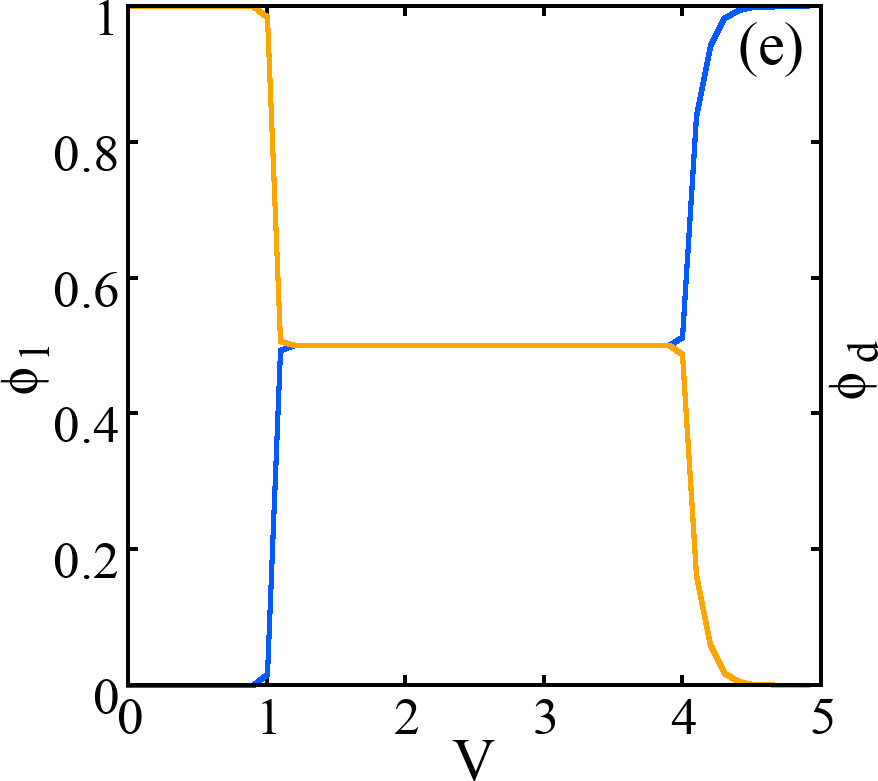}\hspace{0.2cm}
		\includegraphics[width=0.23 \textwidth,height=0.19\textwidth]{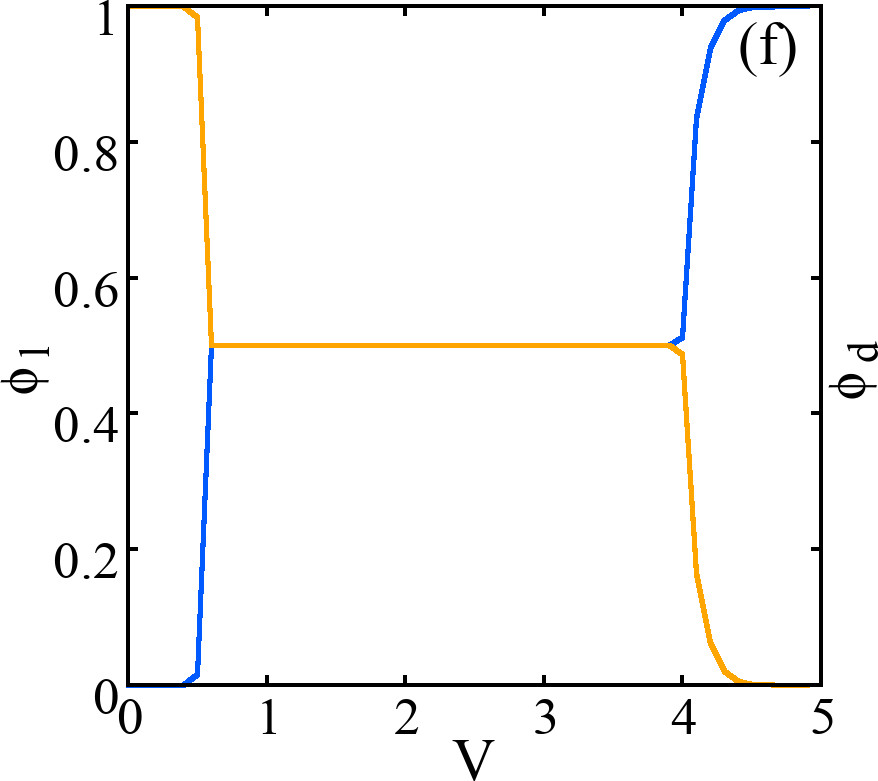}\hspace{0.2cm}
		\includegraphics[width=0.23 \textwidth,height=0.19\textwidth]{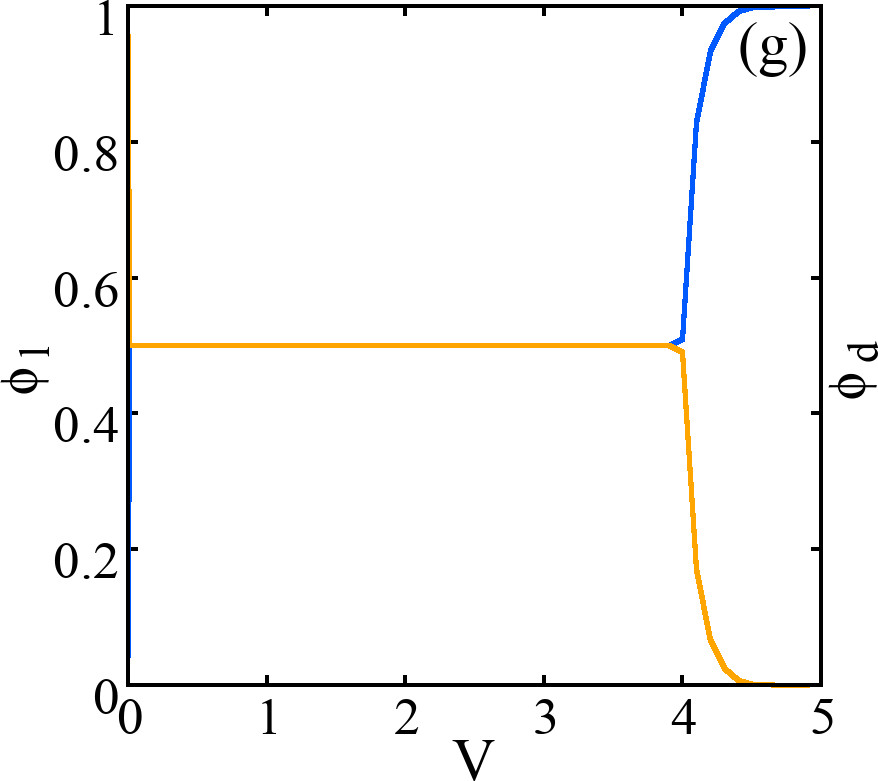}\hspace{0.2cm}
		\includegraphics[width=0.23 \textwidth,height=0.19\textwidth]{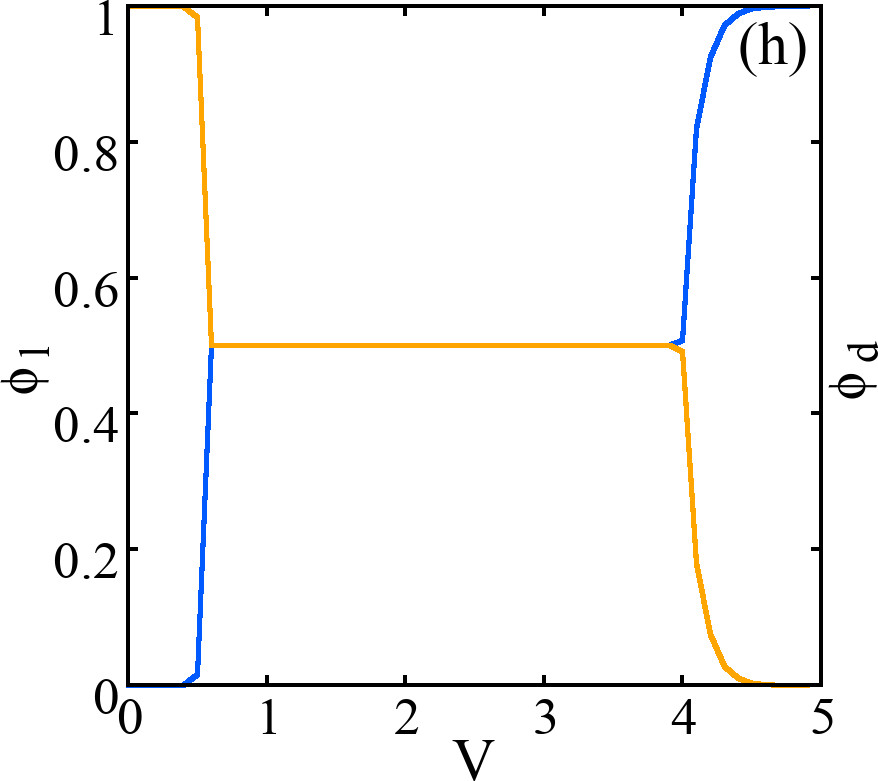}\hspace{0.2cm}
		\caption{The fraction of localized states ($\phi_l$) in blue and delocalized states ($\phi_d$) in yellow corresponding to the parameters of Fig~\ref{Fig:2}. States with $IPR\gtrsim0.1$ was considered localized, otherwise the states are considered to be delocalized in nature.}
		\label{Fig:3}
	\end{figure*}

    We introduce the wave function as,
    \begin{gather}
        {\psi_{n}^{m}}=\begin{pmatrix}
            {\psi_{n,A}^{m}}\\
            {\psi_{n,B}^{m}}
        \end{pmatrix},
        \label{eqn:10}
    \end{gather}
    where, ${\psi_{n,x}^{m}}$ is the normalized wave function of eigenstate labelled by $m$ at site $n$ for the chain $x=A,B$. Substituting Eq.~(\ref{eqn:10}) in Eq.~(\ref{eqn:9}), we obtain,
    \begin{equation}
        \Big(E_{m}\mathbb{1}-\epsilon(n)\Big){\psi_{n}^{m}}=T_1{\psi_{n+1}^{m}}+T_2{\psi_{n-1}^{m}}
        \label{eqn:11}
    \end{equation}
    Eq.~(\ref{eqn:11}) can be disintegrated into the following coupled equations:
    \begin{eqnarray}
       \Big(E_{m}-V_n\Big){\psi_{n,A}^{m}}=t_{L}{\psi_{n+1,A}^{m}}+\nonumber \\t_{R}{\psi_{n-1,A}^{m}}+w_{L}{\psi_{n+1,B}^{m}}+u_{R}{\psi_{n-1,B}^{m}},
        \label{eqn:12}
    \end{eqnarray}
	and
    \begin{eqnarray}
       \Big(E_{m}-V_n\Big){\psi_{n,B}^{m}}=t_{L}{\psi_{n+1,B}^{m}}+\nonumber \\t_{R}{\psi_{n-1,B}^{m}}+w_{R}{\psi_{n-1,A}^{m}}+u_{L}{\psi_{n+1,A}^{m}}.
        \label{eqn:13}
    \end{eqnarray}
    Applying the following canonical transformation 
    \begin{equation}
        {\psi_{n}^{m \pm}}=\frac{{\psi_{n,A}^{m}} \pm {\psi_{n,B}^{m}}}{\sqrt{2}}
    \end{equation}
    and with the following restrictions, i.e., $u_R=w_R=u_1$ and $u_L=w_L=u_2$, the system can be exactly mapped to two uncoupled QHN chains.
    This can be explicitly written as,
    \begin{eqnarray}
       \Big(E_{m}-V_n\Big){\psi_{n}^{m +}}=\Big(t_{L}+u_2\Big){\psi_{n+1}^{m +}}+\Big(t_{R}+u_1 \Big) {\psi_{n-1}^{m+}}~~~~~~
        \label{eqn:14}
    \end{eqnarray}
    and 
    \begin{eqnarray}
       \Big(E_{m}-V_n\Big){\psi_{n}^{m-}}=\Big(t_{L}-u_2\Big){\psi_{n+1}^{m-}}+\Big(t_{R}-u_1 \Big) {\psi_{n-1}^{m-}}~~~~~~
        \label{eqn:15}
    \end{eqnarray}

    The full spectrum is therefore composed of the spectra of the two uncoupled QHN chains, i.e., ${E_{m}^{-}}= E_{m}-V_n$ and $
	{E_{m}^{+}}= E_{m}-V_n$, which are identical. From Eqs.~\ref{eqn:14} and \ref{eqn:15}, one expects two localization transitions at two critical strengths of the quasiperiodic potential at \cite{Jiang},
	\begin{equation}
   	 V_{c1}=2\Big[\text{max}(|t_L-u_2|,|t_R-u_1|)\Big],
    \label{eqn:16}
	\end{equation}
and
\begin{equation}
    V_{c2}=2\Big[\text{max}(|t_L+u_2|,|t_R+u_1|)\Big].
    \label{eqn:17}
\end{equation}

	$V_{c1}$ provides the maximum value of quasiperiodic potential below which all the eigenstates are delocalized. $V_{c2}$ is that strength of the potential above which all the states become completely localized.
	It is interesting to note that one can engineer a system where $V_{c1}$ is zero, when the conditions $|t_L-u_2|=0$ and $|t_R-u_1|=0$ are simultaneously satisfied.

	\section{\label{sec:Results}Numerical Results and Discussions}
	            
	\par In this section, we analyse the phase diagram of the DL transition in the presence of a strong interchain coupling between the two QHN chains A and B.
	The ratio of intrachain hopping strengths of the chains $A(B)$ is $t_L/t_{R}=0.5$.
	In the upper panel of Fig.~\ref{Fig:2}, we consider the case of symmetric interchain coupling.
	It is clear from Figs.~\ref{Fig:2}(a)-(d) that the DL transition does not occur at $V_c=2\text{max}[J_R,J_L]$ (which is the critical value of DL transition in QHN chain).
	It is clearly visible that all the eigenstates are perfectly delocalized for $V\lesssim V_{c1}$  and localized for $V\gtrsim V_{c2}$. The value of $V_{c1}$ and $V_{c2}$ as determined in Eqs.~(\ref{eqn:16}) and (\ref{eqn:17}) agrees excellently with our numerical estimate.
	Furthermore, as is evident, the eigenstates between these two critical points are a mixture of both the delocalized and localized states, separated at a critical energy, termed as the mobility edge.
	\par Next, we consider the case when the interchain coupling is asymmetric in nature in the lower panel of Fig.~\ref{Fig:2}.
	It is clear that the localization behavior drastically changes upon considering a particular strength of asymmetricity, i.e, say, $u_{L}=0.5,~u_{R}=1.0$ and $w_{L}=0.5,~w_{R}=1.0$.
	Such a tendency of $V_{c1}$ approaching 0 is expected when $|t_L-u_2|$ and $|t_R-u_1|$ are both zero, as already explained.
	This particular case is og interest since the localized states appear even for a low value of the quasiperiodic potential, similar to the 1D original Anderson model, although in this case not all the states are localized.
	\par To have a closer look into the nature of states in between $V_{c1}$ and $V_{c2}$, we check the fraction of localized ($\phi_l$) and delocalized states ($\phi_d$).
	We consider the states with $IPR\gtrsim0.1$ as being absolutely localized, and below the limit the states are considered to be delocalized.
	We examine these different regions separately by plotting the fraction of localized and delocalized states as a function of the quasiperiodic potential corresponding to the parameters of Figs.~\ref{Fig:2}. From Figs.~\ref{Fig:3}(a-h), we can easily infer that there is a co-existence of localized and delocalized states for a wide regime in the quasiperiodic potential.
	Moreover, interestingly $50\%$ of these states are delocalized while the remaining states are localized. 
	This proportionate behaviour is consistent throughout the entire intermediate region.
	Furthermore, it is also important to see that in the case where $V_{c1}=0$, exactly $50\%$ localized states appear at even a tiny quasiperiodic potential, as previously discussed.

	 	\begin{figure}[]
		\centering
		\includegraphics[width=0.23\textwidth,height=0.23\textwidth]{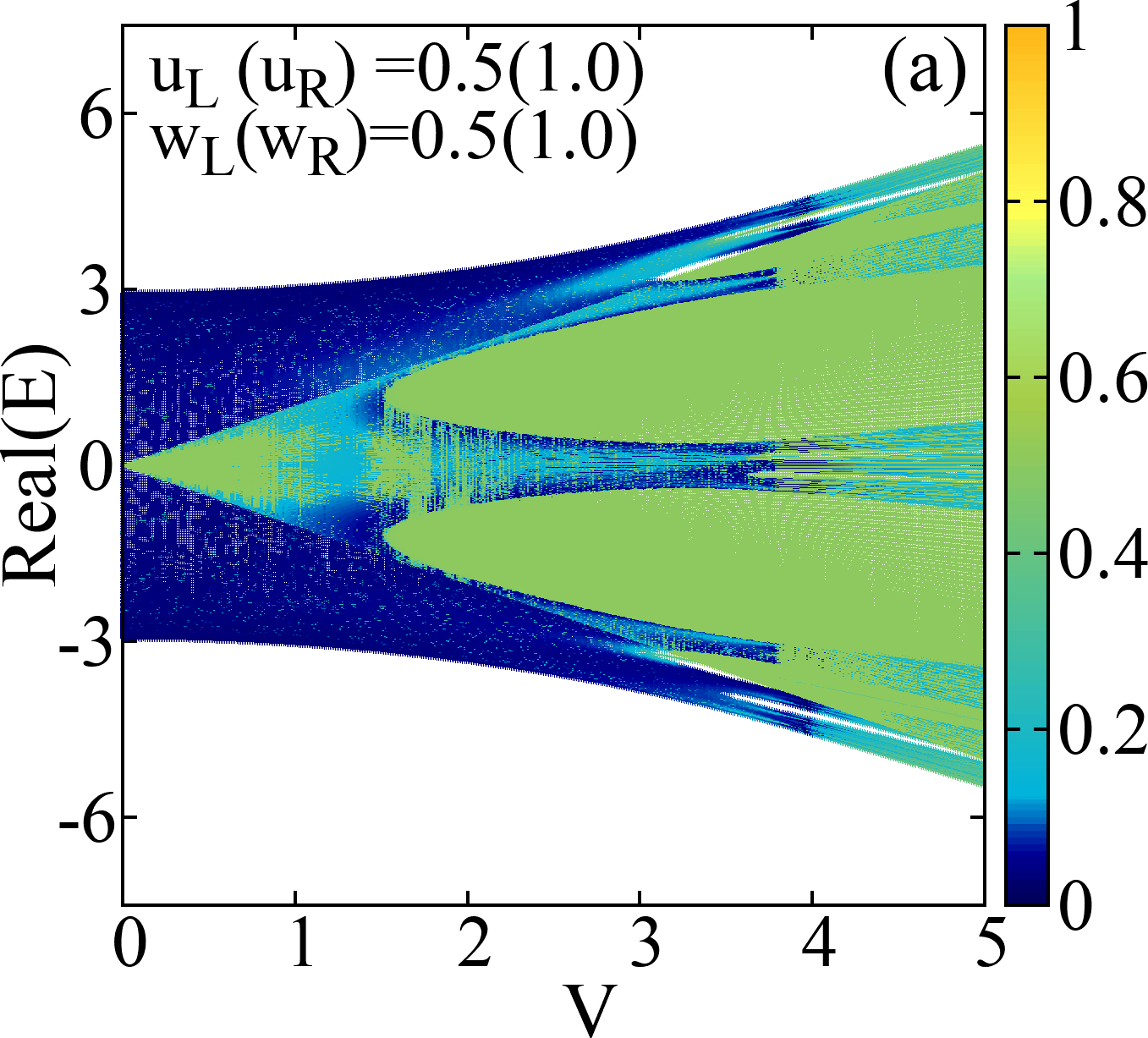}\vspace{0.2cm}
		\includegraphics[width=0.23 \textwidth,height=0.23\textwidth]{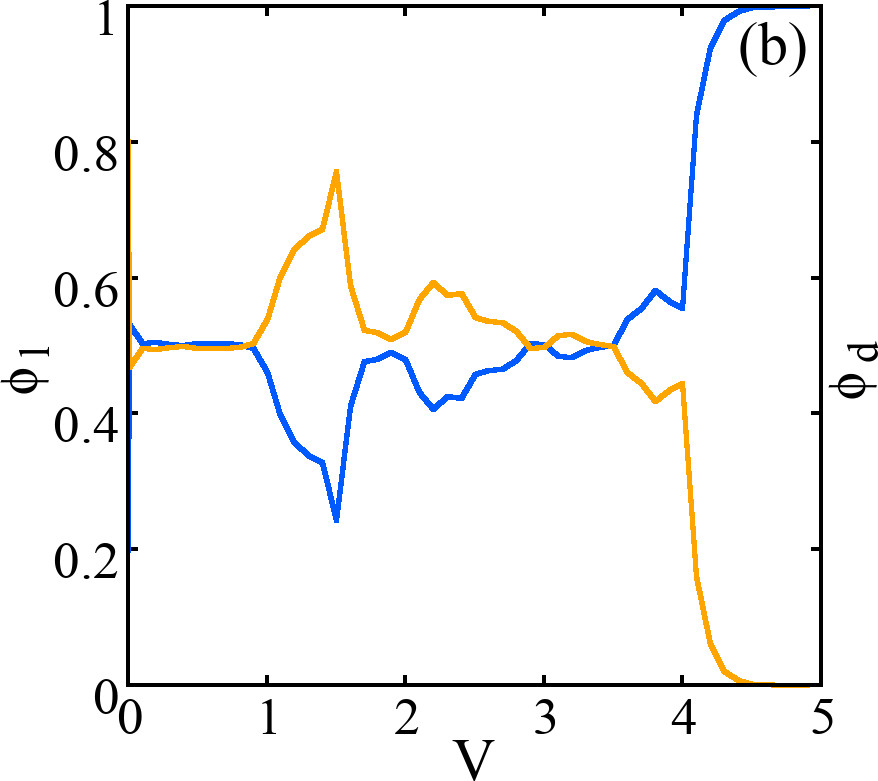}
		\includegraphics[width=0.14 \textwidth,height=0.13\textwidth]{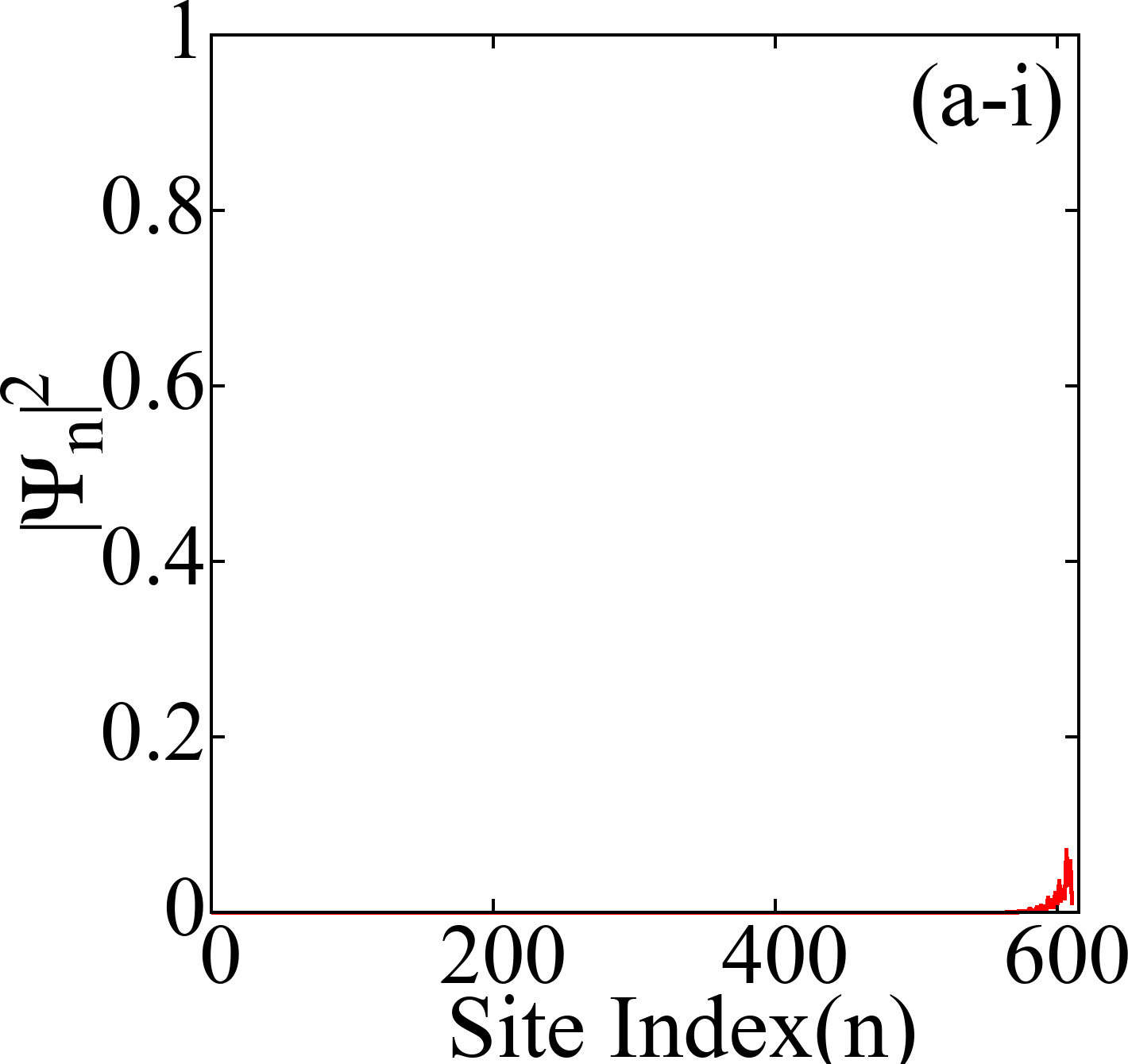}
		\includegraphics[width=0.14 \textwidth,height=0.13\textwidth]{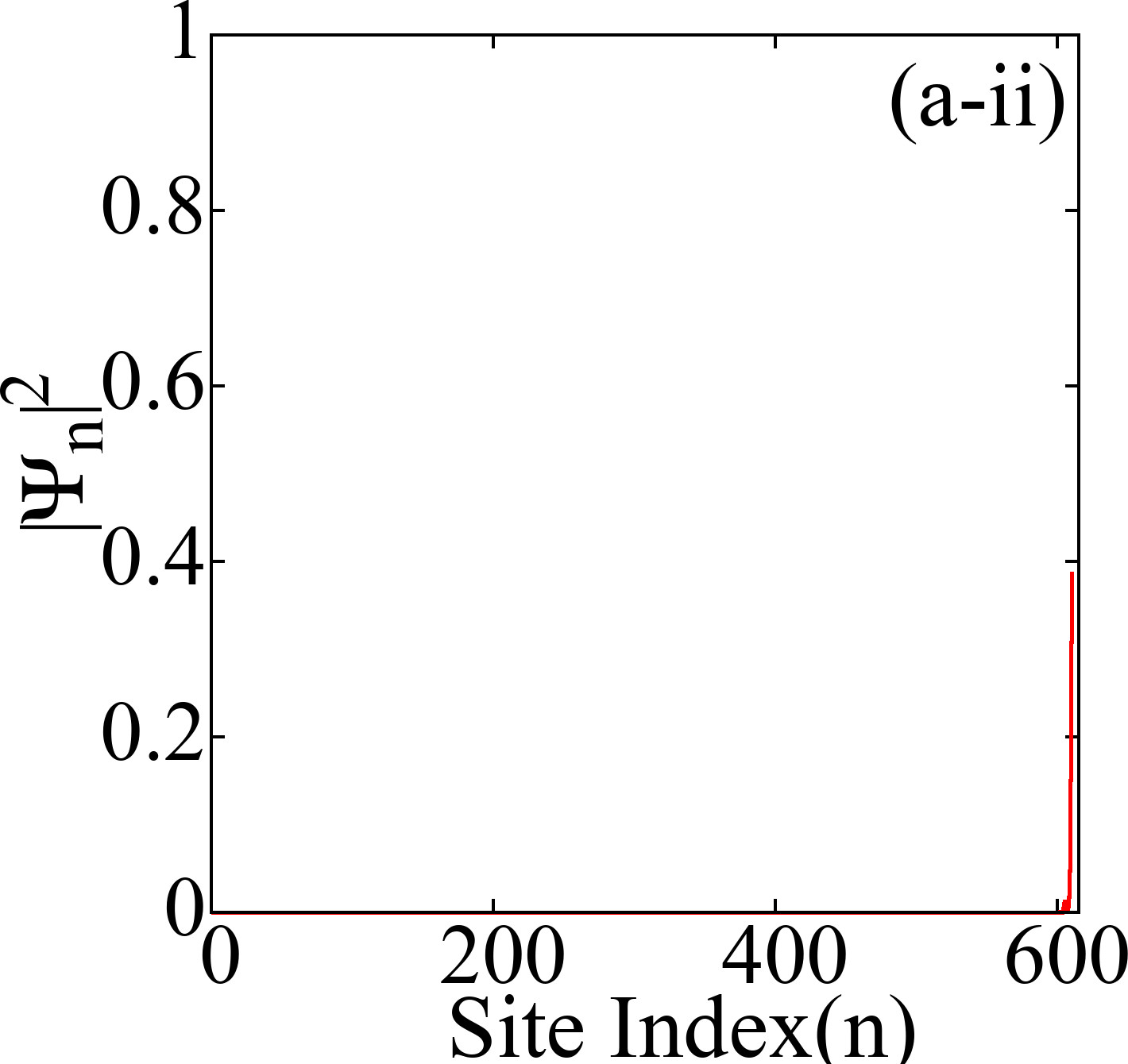}
		\includegraphics[width=0.14 \textwidth,height=0.13\textwidth]{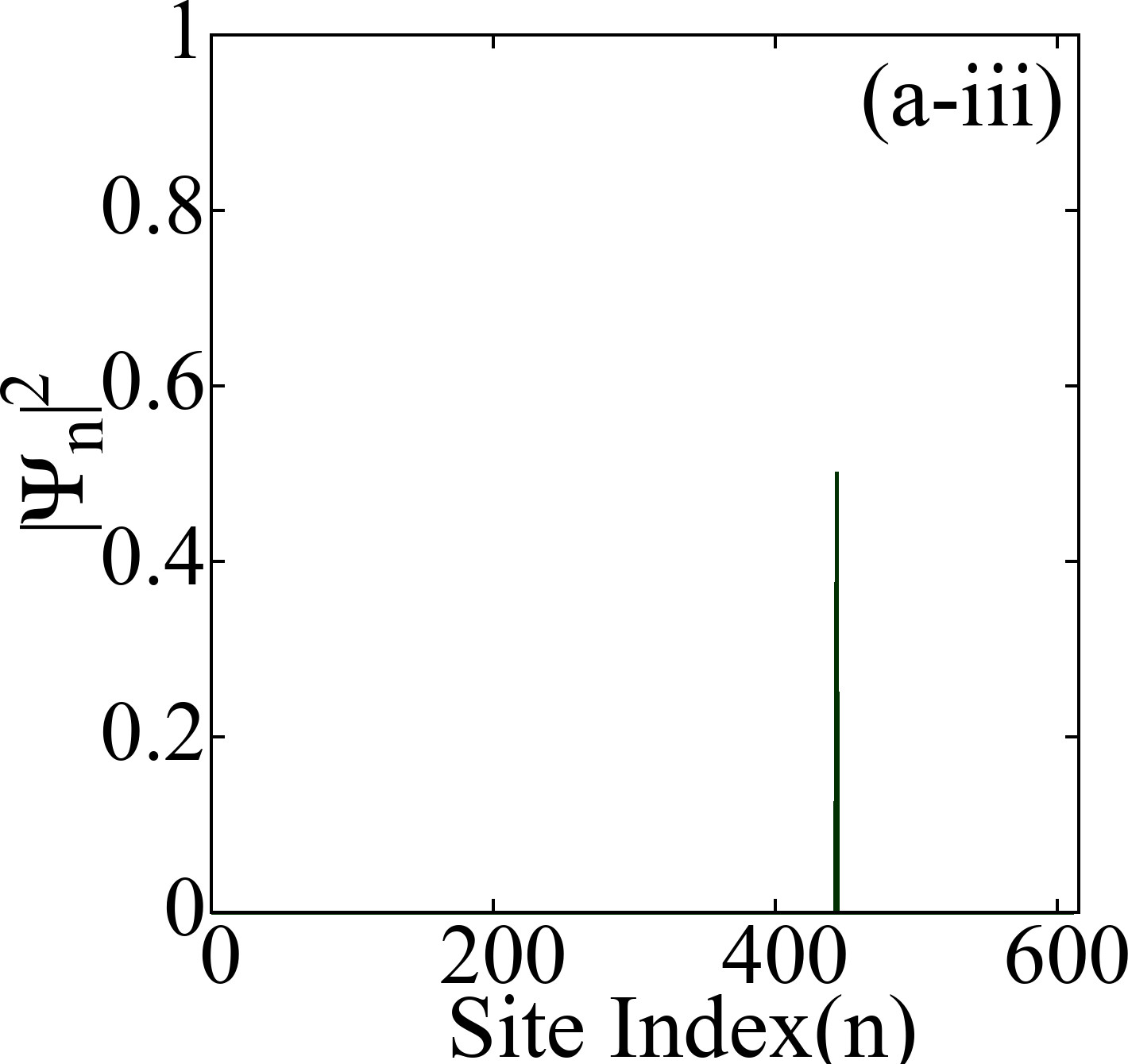}
		\caption{The localization behavior in a strongly coupled QHN Hamiltonian with $t_{L}=0.5,t_{R}=1.0$ for a lattice with $N=610$ sites under OBC with interchain hopping between the two adjacent unit cells. (a) Projection of $IPR$ as a function of the real part of the eigen-energy and quasiperiodic potential ($V$), where the DL transition is shown as a dark blue to green color transition. The other parameters of interchain coupling are: (a)$u_{L}=0.5,u_{R}=1.0,w_{L}=0.5~\text{and}~w_{R}=1.0$. (b) The fraction of localized states ($\phi_l$) in green and delocalized states ($\phi_d$) in yellow, corresponding to the parameters of Fig~\ref{Fig:2}. Figs.~a(i-iii) in the lower panel demonstrates the behavior of the wavefunction probabilities at different latice sites corresponding to the figure in the upper panel at $V=1.5$. (a-i) skin modes (dark blue regime of $IPR$), (a-ii) skin modes in the light blue regime of $IPR$, and (a-iii) localized regime (in green regime of $IPR$).}
		\label{Fig:4}
	\end{figure}
	
  	\par As already elucidated, Fig.~\ref{Fig:2}(g) gives rise to an interesting outcome of localization at a very minute value of the quasiperiodic potential $V$. Therefore, in order to understand whether the same behavior is retained under the OBC, we plot the phase diagram in Fig.~\ref{Fig:4}(a). However, from Fig~\ref{Fig:4}(b), we can infer that proportion of delocalized and localized states does not remain same (i.e., at $50\%$) when the boundaries are open.
  	It is clear, that the localized wavefunctions under PBC become delocalized(skin modes under the OBC) since $\phi_d$ increases. From Fig.~\ref{Fig:4}(a-i) (state picked up from the dark blue regime of the phase diagram), it is clear that the state becomes a skin state under OBC as expected.
  	However, we have found out that the light blue regime infact consists of both skin states (localized at right edge as demonstrated in Fig.~\ref{Fig:4}(a-ii)) and localized states (where the localization is not necessarily towards the right edge as shown in Fig.~\ref{Fig:4}(a-iii).
  	This is in stark contrast to the 1D HN systems in the absence of the coupling.
  	One can therefore infer that additional skin modes are formed from the localized states under OBC due to the coupling between such QHN chains, and hence the one-to-one correspondence between the delocalized(skin) states under PBC(OBC) breaks down in the presence of the coupling. 
	
	\begin{figure}[]
		\centering
		\includegraphics[width=0.35 \textwidth,height=0.24\textwidth]{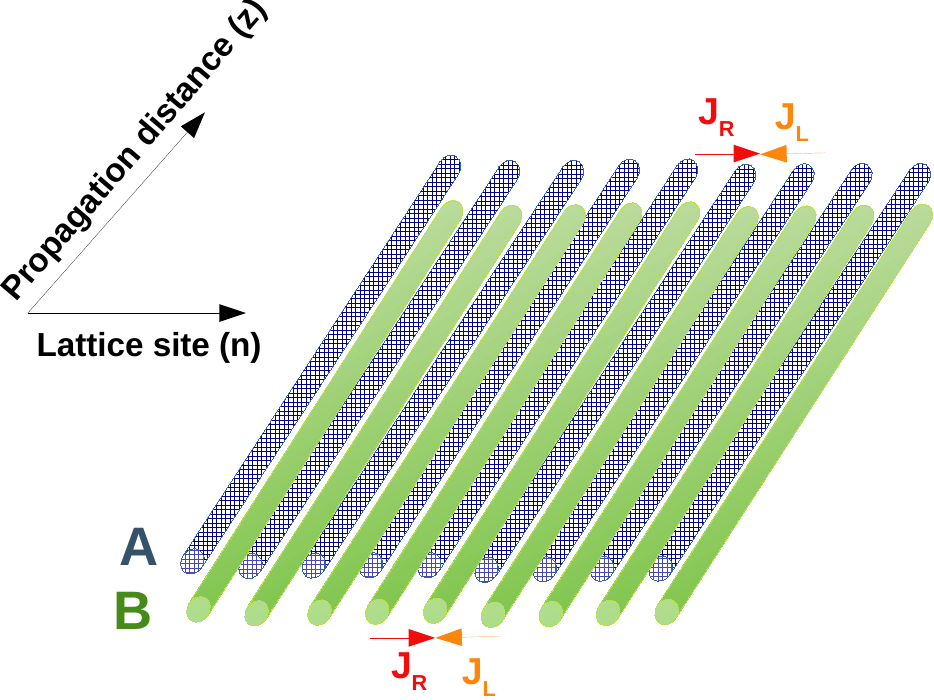}
		\caption{Schematic diagram of the coupled waveguide with asymmetric hopping. Atoms $A$ and $B$ which depict the waveguide channels in the optical set-up are depicted in blue and green respectively.}
		\label{Fig:ER}
	\end{figure}

	\section{\label{sec:Expt}Possible Experimental Implementation in Coupled Waveguides}
	The equation of a coupled waveguide array at position $n$ is written in the form,
	\begin{eqnarray}
		-i\frac{d\psi_n}{dz}=J_L \psi_{n+1} + J_R \psi_{n-1} + V_n \psi_{n}
		\label{Eq:waveguide_eqn}
	\end{eqnarray}
	where $J_L$ and $J_R$ tune the spacing in between the waveguides, and is non-Hermitian in the usual sense. Eq.~\ref{Eq:waveguide_eqn} is an optical analogue of the Schrodinger equation where the time $t$ is replaced by the spatial distance between the parallel waveguides $z$, due to the mathematical equivalence between the two \citep{kokkinakis2024andersonlocalizationversushopping,PhysRevLett.109.106402}. Since we have two atoms (A and B) in a unit cell, we can consider two layers of waveguided arrays (called coupled waveguided arrays) as depicted in the schematic given in Fig.~\ref{Fig:ER}. Such coupled waveguides have already been fabricated on a semiconducting AlGaAs substrate when $J_L=J_R$ \cite{PhysRevLett.109.106402}. The array is composed of a core layer sandwiched between two cladding layers, where the upper cladding layer is etched quasiperiodically, where one can modulate the width of the waveguides quasiperiodically to realize the quasiperiodic onsite potential. The etching	makes the core beneath it have a lower effective refraction index, resulting in a array of coupled 1D waveguides.
	One can tune $J_L$ and $J_R$ using a beam-splitter.
	We consider another coupled waveguide placed exactly below it, which could mimic the coupled QHN system as discussed in our main text.
	Since our work demonstrates the avenue to tune the strengths of $V_{c1}$ and $V_{c2}$ to engineer the localization transitions, such a coupled waveguided array can prove to be a boon to experimentalists working in such optical set-ups.

   	\section{\label{sec:conclusion}Conclusions}  

    To summarize, this work scrutinizes the different localization attributes in non-Hermitian coupled quasiperiodic chains. 
 	The nature of DL transition at a threshold of the quasiperiodic potential ($V_{c}=2$) for NH-AAH chains with parameters $t_{L}=0.5~\text{and}~t_{R}=1.0$ is well-known.
	However, unlike the generic DL transition in NH-AAH chains, a strong coupling between the atoms of adjacent unit cells of the two HN chains possessing the same directionalities under PBC renders an intermediate region, wherein the eigenstates are a mixture of equal proportion of delocalized and localized states.
	Interestingly, for the counterpart with asymmetricity with specific hopping amplitudes, this intermediate region appears even in the 
 	presence of very tiny quasiperiodic potential, where the localized and delocalized states coexist. In this case as well, the proportion of localized and delocalized states remains identical.
 	Moreover, under an OBC, we find a mixture of skin states and localized states in a regime of the localized portion in the PBC phase diagram. This is in contrary to the conventional HN systems where the localized states under OBC can either be skin modes or be completely localized and the usual PBC-OBC correspondence that leads the delocalized states to become skin states, keeping the localized states intact completely breaks down in the presence of the coupling.
 	We believe that these rich phases due to the coupling in non-Hermitian systems can be utilised in experiments related to coupled waveguides.

    \section{Acknowledgements}  
    
	The authors are thankful to the High Performance Computing (HPC) facilities of the National Institute of Technology (Rourkela) and that procured from SERB (DST), India (Grant No. EMR/2015/001227). A.C. acknowledges CSIR-HRDG, India, for providing financial support via File No.- 09/983(0047)/2020-EMR-I.

\bibliography{VarXiv.bib}

\begin{thebibliography}{41}%
\makeatletter
\providecommand \@ifxundefined [1]{%
 \@ifx{#1\undefined}
}%
\providecommand \@ifnum [1]{%
 \ifnum #1\expandafter \@firstoftwo
 \else \expandafter \@secondoftwo
 \fi
}%
\providecommand \@ifx [1]{%
 \ifx #1\expandafter \@firstoftwo
 \else \expandafter \@secondoftwo
 \fi
}%
\providecommand \natexlab [1]{#1}%
\providecommand \enquote  [1]{``#1''}%
\providecommand \bibnamefont  [1]{#1}%
\providecommand \bibfnamefont [1]{#1}%
\providecommand \citenamefont [1]{#1}%
\providecommand \href@noop [0]{\@secondoftwo}%
\providecommand \href [0]{\begingroup \@sanitize@url \@href}%
\providecommand \@href[1]{\@@startlink{#1}\@@href}%
\providecommand \@@href[1]{\endgroup#1\@@endlink}%
\providecommand \@sanitize@url [0]{\catcode `\\12\catcode `\$12\catcode
  `\&12\catcode `\#12\catcode `\^12\catcode `\_12\catcode `\%12\relax}%
\providecommand \@@startlink[1]{}%
\providecommand \@@endlink[0]{}%
\providecommand \url  [0]{\begingroup\@sanitize@url \@url }%
\providecommand \@url [1]{\endgroup\@href {#1}{\urlprefix }}%
\providecommand \urlprefix  [0]{URL }%
\providecommand \Eprint [0]{\href }%
\providecommand \doibase [0]{https://doi.org/}%
\providecommand \selectlanguage [0]{\@gobble}%
\providecommand \bibinfo  [0]{\@secondoftwo}%
\providecommand \bibfield  [0]{\@secondoftwo}%
\providecommand \translation [1]{[#1]}%
\providecommand \BibitemOpen [0]{}%
\providecommand \bibitemStop [0]{}%
\providecommand \bibitemNoStop [0]{.\EOS\space}%
\providecommand \EOS [0]{\spacefactor3000\relax}%
\providecommand \BibitemShut  [1]{\csname bibitem#1\endcsname}%
\let\auto@bib@innerbib\@empty
\bibitem [{\citenamefont {Anderson}(1958)}]{Anderson}%
  \BibitemOpen
  \bibfield  {author} {\bibinfo {author} {\bibfnamefont {P.~W.}\ \bibnamefont
  {Anderson}},\ }\bibfield  {title} {\bibinfo {title} {Absence of diffusion in
  certain random lattices},\ }\href {https://doi.org/10.1103/PhysRev.109.1492}
  {\bibfield  {journal} {\bibinfo  {journal} {Phys. Rev.}\ }\textbf {\bibinfo
  {volume} {109}},\ \bibinfo {pages} {1492} (\bibinfo {year}
  {1958})}\BibitemShut {NoStop}%
\bibitem [{\citenamefont {Katsanos}\ \emph {et~al.}(1998)\citenamefont
  {Katsanos}, \citenamefont {Evangelou},\ and\ \citenamefont
  {Lambert}}]{katsanos1998superconductivity}%
  \BibitemOpen
  \bibfield  {author} {\bibinfo {author} {\bibfnamefont {D.~E.}\ \bibnamefont
  {Katsanos}}, \bibinfo {author} {\bibfnamefont {S.~N.}\ \bibnamefont
  {Evangelou}},\ and\ \bibinfo {author} {\bibfnamefont {C.~J.}\ \bibnamefont
  {Lambert}},\ }\bibfield  {title} {\bibinfo {title} {Superconductivity-induced
  anderson localization},\ }\href {https://doi.org/10.1103/PhysRevB.58.2442}
  {\bibfield  {journal} {\bibinfo  {journal} {Phys. Rev. B}\ }\textbf {\bibinfo
  {volume} {58}},\ \bibinfo {pages} {2442} (\bibinfo {year}
  {1998})}\BibitemShut {NoStop}%
\bibitem [{\citenamefont {Burmistrov}\ \emph {et~al.}(2012)\citenamefont
  {Burmistrov}, \citenamefont {Gornyi},\ and\ \citenamefont
  {Mirlin}}]{burmistrov2012enhancement}%
  \BibitemOpen
  \bibfield  {author} {\bibinfo {author} {\bibfnamefont {I.~S.}\ \bibnamefont
  {Burmistrov}}, \bibinfo {author} {\bibfnamefont {I.~V.}\ \bibnamefont
  {Gornyi}},\ and\ \bibinfo {author} {\bibfnamefont {A.~D.}\ \bibnamefont
  {Mirlin}},\ }\bibfield  {title} {\bibinfo {title} {Enhancement of the
  critical temperature of superconductors by anderson localization},\ }\href
  {https://doi.org/10.1103/PhysRevLett.108.017002} {\bibfield  {journal}
  {\bibinfo  {journal} {Phys. Rev. Lett.}\ }\textbf {\bibinfo {volume} {108}},\
  \bibinfo {pages} {017002} (\bibinfo {year} {2012})}\BibitemShut {NoStop}%
\bibitem [{\citenamefont {Anderson}\ \emph {et~al.}(1983)\citenamefont
  {Anderson}, \citenamefont {Muttalib},\ and\ \citenamefont
  {Ramakrishnan}}]{leavens1985anderson}%
  \BibitemOpen
  \bibfield  {author} {\bibinfo {author} {\bibfnamefont {P.~W.}\ \bibnamefont
  {Anderson}}, \bibinfo {author} {\bibfnamefont {K.~A.}\ \bibnamefont
  {Muttalib}},\ and\ \bibinfo {author} {\bibfnamefont {T.~V.}\ \bibnamefont
  {Ramakrishnan}},\ }\bibfield  {title} {\bibinfo {title} {Theory of the
  "universal" degradation of tc in high-temperature superconductors},\ }\href
  {https://doi.org/10.1103/PhysRevB.28.117} {\bibfield  {journal} {\bibinfo
  {journal} {Phys. Rev. B}\ }\textbf {\bibinfo {volume} {28}},\ \bibinfo
  {pages} {117} (\bibinfo {year} {1983})}\BibitemShut {NoStop}%
\bibitem [{\citenamefont {Sarychev}\ \emph {et~al.}(1999)\citenamefont
  {Sarychev}, \citenamefont {Shubin},\ and\ \citenamefont
  {Shalaev}}]{sarychev1999anderson}%
  \BibitemOpen
  \bibfield  {author} {\bibinfo {author} {\bibfnamefont {A.~K.}\ \bibnamefont
  {Sarychev}}, \bibinfo {author} {\bibfnamefont {V.~A.}\ \bibnamefont
  {Shubin}},\ and\ \bibinfo {author} {\bibfnamefont {V.~M.}\ \bibnamefont
  {Shalaev}},\ }\bibfield  {title} {\bibinfo {title} {Anderson localization of
  surface plasmons and nonlinear optics of metal-dielectric composites},\
  }\href {https://doi.org/10.1103/PhysRevB.60.16389} {\bibfield  {journal}
  {\bibinfo  {journal} {Phys. Rev. B}\ }\textbf {\bibinfo {volume} {60}},\
  \bibinfo {pages} {16389} (\bibinfo {year} {1999})}\BibitemShut {NoStop}%
\bibitem [{\citenamefont {Lahini}\ \emph {et~al.}(2008)\citenamefont {Lahini},
  \citenamefont {Avidan}, \citenamefont {Pozzi}, \citenamefont {Sorel},
  \citenamefont {Morandotti}, \citenamefont {Christodoulides},\ and\
  \citenamefont {Silberberg}}]{lahini2008anderson}%
  \BibitemOpen
  \bibfield  {author} {\bibinfo {author} {\bibfnamefont {Y.}~\bibnamefont
  {Lahini}}, \bibinfo {author} {\bibfnamefont {A.}~\bibnamefont {Avidan}},
  \bibinfo {author} {\bibfnamefont {F.}~\bibnamefont {Pozzi}}, \bibinfo
  {author} {\bibfnamefont {M.}~\bibnamefont {Sorel}}, \bibinfo {author}
  {\bibfnamefont {R.}~\bibnamefont {Morandotti}}, \bibinfo {author}
  {\bibfnamefont {D.~N.}\ \bibnamefont {Christodoulides}},\ and\ \bibinfo
  {author} {\bibfnamefont {Y.}~\bibnamefont {Silberberg}},\ }\bibfield  {title}
  {\bibinfo {title} {Anderson localization and nonlinearity in one-dimensional
  disordered photonic lattices},\ }\href
  {https://doi.org/10.1103/PhysRevLett.100.013906} {\bibfield  {journal}
  {\bibinfo  {journal} {Phys. Rev. Lett.}\ }\textbf {\bibinfo {volume} {100}},\
  \bibinfo {pages} {013906} (\bibinfo {year} {2008})}\BibitemShut {NoStop}%
\bibitem [{\citenamefont {Jovi\ifmmode~\acute{c}\else \'{c}\fi{}}\ \emph
  {et~al.}(2011)\citenamefont {Jovi\ifmmode~\acute{c}\else \'{c}\fi{}},
  \citenamefont {Kivshar}, \citenamefont {Denz},\ and\ \citenamefont
  {Beli\ifmmode~\acute{c}\else \'{c}\fi{}}}]{jovic2011anderson}%
  \BibitemOpen
  \bibfield  {author} {\bibinfo {author} {\bibfnamefont {D.~M.}\ \bibnamefont
  {Jovi\ifmmode~\acute{c}\else \'{c}\fi{}}}, \bibinfo {author} {\bibfnamefont
  {Y.~S.}\ \bibnamefont {Kivshar}}, \bibinfo {author} {\bibfnamefont
  {C.}~\bibnamefont {Denz}},\ and\ \bibinfo {author} {\bibfnamefont {M.~R.}\
  \bibnamefont {Beli\ifmmode~\acute{c}\else \'{c}\fi{}}},\ }\bibfield  {title}
  {\bibinfo {title} {Anderson localization of light near boundaries of
  disordered photonic lattices},\ }\href
  {https://doi.org/10.1103/PhysRevA.83.033813} {\bibfield  {journal} {\bibinfo
  {journal} {Phys. Rev. A}\ }\textbf {\bibinfo {volume} {83}},\ \bibinfo
  {pages} {033813} (\bibinfo {year} {2011})}\BibitemShut {NoStop}%
\bibitem [{\citenamefont {Qiao}\ \emph {et~al.}(2019)\citenamefont {Qiao},
  \citenamefont {Ye}, \citenamefont {Zheng},\ and\ \citenamefont
  {Chen}}]{qiao2019cavity}%
  \BibitemOpen
  \bibfield  {author} {\bibinfo {author} {\bibfnamefont {Y.}~\bibnamefont
  {Qiao}}, \bibinfo {author} {\bibfnamefont {F.}~\bibnamefont {Ye}}, \bibinfo
  {author} {\bibfnamefont {Y.}~\bibnamefont {Zheng}},\ and\ \bibinfo {author}
  {\bibfnamefont {X.}~\bibnamefont {Chen}},\ }\bibfield  {title} {\bibinfo
  {title} {Cavity-enhanced second-harmonic generation in strongly scattering
  nonlinear media},\ }\href {https://doi.org/10.1103/PhysRevA.99.043844}
  {\bibfield  {journal} {\bibinfo  {journal} {Phys. Rev. A}\ }\textbf {\bibinfo
  {volume} {99}},\ \bibinfo {pages} {043844} (\bibinfo {year}
  {2019})}\BibitemShut {NoStop}%
\bibitem [{\citenamefont {Condat}\ and\ \citenamefont
  {Kirkpatrick}(1987)}]{condat1987observability}%
  \BibitemOpen
  \bibfield  {author} {\bibinfo {author} {\bibfnamefont {C.~A.}\ \bibnamefont
  {Condat}}\ and\ \bibinfo {author} {\bibfnamefont {T.~R.}\ \bibnamefont
  {Kirkpatrick}},\ }\bibfield  {title} {\bibinfo {title} {Observability of
  acoustical and optical localization},\ }\href
  {https://doi.org/10.1103/PhysRevLett.58.226} {\bibfield  {journal} {\bibinfo
  {journal} {Phys. Rev. Lett.}\ }\textbf {\bibinfo {volume} {58}},\ \bibinfo
  {pages} {226} (\bibinfo {year} {1987})}\BibitemShut {NoStop}%
\bibitem [{\citenamefont {Cohen}\ \emph {et~al.}(1987)\citenamefont {Cohen},
  \citenamefont {Machta}, \citenamefont {Kirkpatrick},\ and\ \citenamefont
  {Condat}}]{cohen1987crossover}%
  \BibitemOpen
  \bibfield  {author} {\bibinfo {author} {\bibfnamefont {S.~M.}\ \bibnamefont
  {Cohen}}, \bibinfo {author} {\bibfnamefont {J.}~\bibnamefont {Machta}},
  \bibinfo {author} {\bibfnamefont {T.~R.}\ \bibnamefont {Kirkpatrick}},\ and\
  \bibinfo {author} {\bibfnamefont {C.~A.}\ \bibnamefont {Condat}},\ }\bibfield
   {title} {\bibinfo {title} {Crossover in the anderson transition: Acoustic
  localization with a flow},\ }\href
  {https://doi.org/10.1103/PhysRevLett.58.785} {\bibfield  {journal} {\bibinfo
  {journal} {Phys. Rev. Lett.}\ }\textbf {\bibinfo {volume} {58}},\ \bibinfo
  {pages} {785} (\bibinfo {year} {1987})}\BibitemShut {NoStop}%
\bibitem [{\citenamefont {Abrahams}\ \emph {et~al.}(1979)\citenamefont
  {Abrahams}, \citenamefont {Anderson}, \citenamefont {Licciardello},\ and\
  \citenamefont {Ramakrishnan}}]{abrahams1979scaling}%
  \BibitemOpen
  \bibfield  {author} {\bibinfo {author} {\bibfnamefont {E.}~\bibnamefont
  {Abrahams}}, \bibinfo {author} {\bibfnamefont {P.~W.}\ \bibnamefont
  {Anderson}}, \bibinfo {author} {\bibfnamefont {D.~C.}\ \bibnamefont
  {Licciardello}},\ and\ \bibinfo {author} {\bibfnamefont {T.~V.}\ \bibnamefont
  {Ramakrishnan}},\ }\bibfield  {title} {\bibinfo {title} {Scaling theory of
  localization: Absence of quantum diffusion in two dimensions},\ }\href
  {https://doi.org/10.1103/PhysRevLett.42.673} {\bibfield  {journal} {\bibinfo
  {journal} {Phys. Rev. Lett.}\ }\textbf {\bibinfo {volume} {42}},\ \bibinfo
  {pages} {673} (\bibinfo {year} {1979})}\BibitemShut {NoStop}%
\bibitem [{\citenamefont {Aubry}\ and\ \citenamefont
  {Andr{\'e}}(1980)}]{aubry1980analyticity}%
  \BibitemOpen
  \bibfield  {author} {\bibinfo {author} {\bibfnamefont {S.}~\bibnamefont
  {Aubry}}\ and\ \bibinfo {author} {\bibfnamefont {G.}~\bibnamefont
  {Andr{\'e}}},\ }\bibfield  {title} {\bibinfo {title} {Analyticity breaking
  and anderson localization in incommensurate lattices},\ }\href
  {https://www.researchgate.net/publication/265502988_Analyticity_breaking_and_Anderson_localization_in_incommensurate_lattices}
  {\bibfield  {journal} {\bibinfo  {journal} {Ann. Israel Phys. Soc}\ }\textbf
  {\bibinfo {volume} {3}},\ \bibinfo {pages} {18} (\bibinfo {year}
  {1980})}\BibitemShut {NoStop}%
\bibitem [{\citenamefont {Harper}(1955)}]{harper1955single}%
  \BibitemOpen
  \bibfield  {author} {\bibinfo {author} {\bibfnamefont {P.~G.}\ \bibnamefont
  {Harper}},\ }\bibfield  {title} {\bibinfo {title} {Single band motion of
  conduction electrons in a uniform magnetic field},\ }\href
  {https://doi.org/10.1088/0370-1298/68/10/304} {\bibfield  {journal} {\bibinfo
   {journal} {Proceedings of the Physical Society. Section A}\ }\textbf
  {\bibinfo {volume} {68}},\ \bibinfo {pages} {874} (\bibinfo {year}
  {1955})}\BibitemShut {NoStop}%
\bibitem [{\citenamefont {Longhi}(2019)}]{longhi2019metal}%
  \BibitemOpen
  \bibfield  {author} {\bibinfo {author} {\bibfnamefont {S.}~\bibnamefont
  {Longhi}},\ }\bibfield  {title} {\bibinfo {title} {Metal-insulator phase
  transition in a non-hermitian aubry-andr\'e-harper model},\ }\href
  {https://doi.org/10.1103/PhysRevB.100.125157} {\bibfield  {journal} {\bibinfo
   {journal} {Phys. Rev. B}\ }\textbf {\bibinfo {volume} {100}},\ \bibinfo
  {pages} {125157} (\bibinfo {year} {2019})}\BibitemShut {NoStop}%
\bibitem [{\citenamefont {Tong}(1994)}]{tong1994localization}%
  \BibitemOpen
  \bibfield  {author} {\bibinfo {author} {\bibfnamefont {P.}~\bibnamefont
  {Tong}},\ }\bibfield  {title} {\bibinfo {title} {Localization and mobility
  edges in one-dimensional deterministic potentials},\ }\href
  {https://doi.org/10.1103/PhysRevB.50.11318} {\bibfield  {journal} {\bibinfo
  {journal} {Phys. Rev. B}\ }\textbf {\bibinfo {volume} {50}},\ \bibinfo
  {pages} {11318} (\bibinfo {year} {1994})}\BibitemShut {NoStop}%
\bibitem [{\citenamefont {Ossipov}\ \emph {et~al.}(2001)\citenamefont
  {Ossipov}, \citenamefont {Weiss}, \citenamefont {Kottos},\ and\ \citenamefont
  {Geisel}}]{ossipov2001quantum}%
  \BibitemOpen
  \bibfield  {author} {\bibinfo {author} {\bibfnamefont {A.}~\bibnamefont
  {Ossipov}}, \bibinfo {author} {\bibfnamefont {M.}~\bibnamefont {Weiss}},
  \bibinfo {author} {\bibfnamefont {T.}~\bibnamefont {Kottos}},\ and\ \bibinfo
  {author} {\bibfnamefont {T.}~\bibnamefont {Geisel}},\ }\bibfield  {title}
  {\bibinfo {title} {Quantum mechanical relaxation of open quasiperiodic
  systems},\ }\href {https://doi.org/10.1103/PhysRevB.64.224210} {\bibfield
  {journal} {\bibinfo  {journal} {Phys. Rev. B}\ }\textbf {\bibinfo {volume}
  {64}},\ \bibinfo {pages} {224210} (\bibinfo {year} {2001})}\BibitemShut
  {NoStop}%
\bibitem [{\citenamefont {Yoo}\ \emph {et~al.}(2020)\citenamefont {Yoo},
  \citenamefont {Lee},\ and\ \citenamefont {Swingle}}]{yoo2020nonequilibrium}%
  \BibitemOpen
  \bibfield  {author} {\bibinfo {author} {\bibfnamefont {Y.}~\bibnamefont
  {Yoo}}, \bibinfo {author} {\bibfnamefont {J.}~\bibnamefont {Lee}},\ and\
  \bibinfo {author} {\bibfnamefont {B.}~\bibnamefont {Swingle}},\ }\bibfield
  {title} {\bibinfo {title} {Nonequilibrium steady state phases of the
  interacting aubry-andr\'e-harper model},\ }\href
  {https://doi.org/10.1103/PhysRevB.102.195142} {\bibfield  {journal} {\bibinfo
   {journal} {Phys. Rev. B}\ }\textbf {\bibinfo {volume} {102}},\ \bibinfo
  {pages} {195142} (\bibinfo {year} {2020})}\BibitemShut {NoStop}%
\bibitem [{\citenamefont {Sokoloff}(1981)}]{Sokoloff}%
  \BibitemOpen
  \bibfield  {author} {\bibinfo {author} {\bibfnamefont {J.~B.}\ \bibnamefont
  {Sokoloff}},\ }\bibfield  {title} {\bibinfo {title} {Band structure and
  localization in incommensurate lattice potentials},\ }\href
  {https://doi.org/10.1103/PhysRevB.23.6422} {\bibfield  {journal} {\bibinfo
  {journal} {Phys. Rev. B}\ }\textbf {\bibinfo {volume} {23}},\ \bibinfo
  {pages} {6422} (\bibinfo {year} {1981})}\BibitemShut {NoStop}%
\bibitem [{\citenamefont {Sokoloff}(1985)}]{sokoloff1985unusual}%
  \BibitemOpen
  \bibfield  {author} {\bibinfo {author} {\bibfnamefont {J.}~\bibnamefont
  {Sokoloff}},\ }\bibfield  {title} {\bibinfo {title} {Unusual band structure,
  wave functions and electrical conductance in crystals with incommensurate
  periodic potentials},\ }\href {https://doi.org/10.1016/0370-1573(85)90088-2}
  {\bibfield  {journal} {\bibinfo  {journal} {Physics Reports}\ }\textbf
  {\bibinfo {volume} {126}},\ \bibinfo {pages} {189} (\bibinfo {year}
  {1985})}\BibitemShut {NoStop}%
\bibitem [{\citenamefont {Kohmoto}(1983)}]{kohmoto1983metal}%
  \BibitemOpen
  \bibfield  {author} {\bibinfo {author} {\bibfnamefont {M.}~\bibnamefont
  {Kohmoto}},\ }\bibfield  {title} {\bibinfo {title} {Metal-insulator
  transition and scaling for incommensurate systems},\ }\href
  {https://doi.org/10.1103/PhysRevLett.51.1198} {\bibfield  {journal} {\bibinfo
   {journal} {Phys. Rev. Lett.}\ }\textbf {\bibinfo {volume} {51}},\ \bibinfo
  {pages} {1198} (\bibinfo {year} {1983})}\BibitemShut {NoStop}%
\bibitem [{\citenamefont {Thouless}(1983)}]{Thouless}%
  \BibitemOpen
  \bibfield  {author} {\bibinfo {author} {\bibfnamefont {D.~J.}\ \bibnamefont
  {Thouless}},\ }\bibfield  {title} {\bibinfo {title} {Bandwidths for a
  quasiperiodic tight-binding model},\ }\href
  {https://doi.org/10.1103/PhysRevB.28.4272} {\bibfield  {journal} {\bibinfo
  {journal} {Phys. Rev. B}\ }\textbf {\bibinfo {volume} {28}},\ \bibinfo
  {pages} {4272} (\bibinfo {year} {1983})}\BibitemShut {NoStop}%
\bibitem [{\citenamefont {Hiramoto}\ and\ \citenamefont
  {Kohmoto}(1992)}]{hiramoto1992electronic}%
  \BibitemOpen
  \bibfield  {author} {\bibinfo {author} {\bibfnamefont {H.}~\bibnamefont
  {Hiramoto}}\ and\ \bibinfo {author} {\bibfnamefont {M.}~\bibnamefont
  {Kohmoto}},\ }\bibfield  {title} {\bibinfo {title} {Electronic spectral and
  wavefunction properties of one-dimensional quasiperiodic systems: A scaling
  approach},\ }\href {https://doi.org/10.1142/S0217979292000153} {\bibfield
  {journal} {\bibinfo  {journal} {International Journal of Modern Physics B}\
  }\textbf {\bibinfo {volume} {06}},\ \bibinfo {pages} {281} (\bibinfo {year}
  {1992})}\BibitemShut {NoStop}%
\bibitem [{\citenamefont {Kramer}\ and\ \citenamefont
  {MacKinnon}(1993)}]{kramer1993localization}%
  \BibitemOpen
  \bibfield  {author} {\bibinfo {author} {\bibfnamefont {B.}~\bibnamefont
  {Kramer}}\ and\ \bibinfo {author} {\bibfnamefont {A.}~\bibnamefont
  {MacKinnon}},\ }\bibfield  {title} {\bibinfo {title} {Localization: theory
  and experiment},\ }\href {https://doi.org/10.1088/0034-4885/56/12/001}
  {\bibfield  {journal} {\bibinfo  {journal} {Reports on Progress in Physics}\
  }\textbf {\bibinfo {volume} {56}},\ \bibinfo {pages} {1469} (\bibinfo {year}
  {1993})}\BibitemShut {NoStop}%
\bibitem [{\citenamefont {Chang}\ \emph {et~al.}(1997)\citenamefont {Chang},
  \citenamefont {Ikezawa},\ and\ \citenamefont
  {Kohmoto}}]{chang1997multifractal}%
  \BibitemOpen
  \bibfield  {author} {\bibinfo {author} {\bibfnamefont {I.}~\bibnamefont
  {Chang}}, \bibinfo {author} {\bibfnamefont {K.}~\bibnamefont {Ikezawa}},\
  and\ \bibinfo {author} {\bibfnamefont {M.}~\bibnamefont {Kohmoto}},\
  }\bibfield  {title} {\bibinfo {title} {Multifractal properties of the wave
  functions of the square-lattice tight-binding model with
  next-nearest-neighbor hopping in a magnetic field},\ }\href
  {https://doi.org/10.1103/PhysRevB.55.12971} {\bibfield  {journal} {\bibinfo
  {journal} {Phys. Rev. B}\ }\textbf {\bibinfo {volume} {55}},\ \bibinfo
  {pages} {12971} (\bibinfo {year} {1997})}\BibitemShut {NoStop}%
\bibitem [{\citenamefont {Roati}\ \emph {et~al.}(2008)\citenamefont {Roati},
  \citenamefont {D’Errico}, \citenamefont {Fallani}, \citenamefont {Fattori},
  \citenamefont {Fort}, \citenamefont {Zaccanti}, \citenamefont {Modugno},
  \citenamefont {Modugno},\ and\ \citenamefont {Inguscio}}]{roati2008anderson}%
  \BibitemOpen
  \bibfield  {author} {\bibinfo {author} {\bibfnamefont {G.}~\bibnamefont
  {Roati}}, \bibinfo {author} {\bibfnamefont {C.}~\bibnamefont {D’Errico}},
  \bibinfo {author} {\bibfnamefont {L.}~\bibnamefont {Fallani}}, \bibinfo
  {author} {\bibfnamefont {M.}~\bibnamefont {Fattori}}, \bibinfo {author}
  {\bibfnamefont {C.}~\bibnamefont {Fort}}, \bibinfo {author} {\bibfnamefont
  {M.}~\bibnamefont {Zaccanti}}, \bibinfo {author} {\bibfnamefont
  {G.}~\bibnamefont {Modugno}}, \bibinfo {author} {\bibfnamefont
  {M.}~\bibnamefont {Modugno}},\ and\ \bibinfo {author} {\bibfnamefont
  {M.}~\bibnamefont {Inguscio}},\ }\bibfield  {title} {\bibinfo {title}
  {Anderson localization of a non-interacting bose-einstein condensate},\
  }\href {https://doi.org/10.1038/nature07071} {\bibfield  {journal} {\bibinfo
  {journal} {Nature}\ }\textbf {\bibinfo {volume} {453}},\ \bibinfo {pages}
  {895} (\bibinfo {year} {2008})}\BibitemShut {NoStop}%
\bibitem [{\citenamefont {Deissler}\ \emph {et~al.}(2010)\citenamefont
  {Deissler}, \citenamefont {Zaccanti}, \citenamefont {Roati}, \citenamefont
  {D’Errico}, \citenamefont {Fattori}, \citenamefont {Modugno}, \citenamefont
  {Modugno},\ and\ \citenamefont {Inguscio}}]{deissler2010delocalization}%
  \BibitemOpen
  \bibfield  {author} {\bibinfo {author} {\bibfnamefont {B.}~\bibnamefont
  {Deissler}}, \bibinfo {author} {\bibfnamefont {M.}~\bibnamefont {Zaccanti}},
  \bibinfo {author} {\bibfnamefont {G.}~\bibnamefont {Roati}}, \bibinfo
  {author} {\bibfnamefont {C.}~\bibnamefont {D’Errico}}, \bibinfo {author}
  {\bibfnamefont {M.}~\bibnamefont {Fattori}}, \bibinfo {author} {\bibfnamefont
  {M.}~\bibnamefont {Modugno}}, \bibinfo {author} {\bibfnamefont
  {G.}~\bibnamefont {Modugno}},\ and\ \bibinfo {author} {\bibfnamefont
  {M.}~\bibnamefont {Inguscio}},\ }\bibfield  {title} {\bibinfo {title}
  {Delocalization of a disordered bosonic system by repulsive interactions},\
  }\href {https://doi.org/10.1038/nphys1635} {\bibfield  {journal} {\bibinfo
  {journal} {Nature physics}\ }\textbf {\bibinfo {volume} {6}},\ \bibinfo
  {pages} {354} (\bibinfo {year} {2010})}\BibitemShut {NoStop}%
\bibitem [{\citenamefont {Schreiber}\ \emph {et~al.}(2015)\citenamefont
  {Schreiber}, \citenamefont {Hodgman}, \citenamefont {Bordia}, \citenamefont
  {Lüschen}, \citenamefont {Fischer}, \citenamefont {Vosk}, \citenamefont
  {Altman}, \citenamefont {Schneider},\ and\ \citenamefont
  {Bloch}}]{schreiber2015observation}%
  \BibitemOpen
  \bibfield  {author} {\bibinfo {author} {\bibfnamefont {M.}~\bibnamefont
  {Schreiber}}, \bibinfo {author} {\bibfnamefont {S.~S.}\ \bibnamefont
  {Hodgman}}, \bibinfo {author} {\bibfnamefont {P.}~\bibnamefont {Bordia}},
  \bibinfo {author} {\bibfnamefont {H.~P.}\ \bibnamefont {Lüschen}}, \bibinfo
  {author} {\bibfnamefont {M.~H.}\ \bibnamefont {Fischer}}, \bibinfo {author}
  {\bibfnamefont {R.}~\bibnamefont {Vosk}}, \bibinfo {author} {\bibfnamefont
  {E.}~\bibnamefont {Altman}}, \bibinfo {author} {\bibfnamefont
  {U.}~\bibnamefont {Schneider}},\ and\ \bibinfo {author} {\bibfnamefont
  {I.}~\bibnamefont {Bloch}},\ }\bibfield  {title} {\bibinfo {title}
  {Observation of many-body localization of interacting fermions in a
  quasirandom optical lattice},\ }\href
  {https://doi.org/10.1126/science.aaa7432} {\bibfield  {journal} {\bibinfo
  {journal} {Science}\ }\textbf {\bibinfo {volume} {349}},\ \bibinfo {pages}
  {842} (\bibinfo {year} {2015})}\BibitemShut {NoStop}%
\bibitem [{\citenamefont {Tang}\ \emph {et~al.}(2021)\citenamefont {Tang},
  \citenamefont {Zhang}, \citenamefont {Zhang},\ and\ \citenamefont
  {Zhang}}]{tang2021localization}%
  \BibitemOpen
  \bibfield  {author} {\bibinfo {author} {\bibfnamefont {L.-Z.}\ \bibnamefont
  {Tang}}, \bibinfo {author} {\bibfnamefont {G.-Q.}\ \bibnamefont {Zhang}},
  \bibinfo {author} {\bibfnamefont {L.-F.}\ \bibnamefont {Zhang}},\ and\
  \bibinfo {author} {\bibfnamefont {D.-W.}\ \bibnamefont {Zhang}},\ }\bibfield
  {title} {\bibinfo {title} {Localization and topological transitions in
  non-hermitian quasiperiodic lattices},\ }\href
  {https://doi.org/10.1103/PhysRevA.103.033325} {\bibfield  {journal} {\bibinfo
   {journal} {Phys. Rev. A}\ }\textbf {\bibinfo {volume} {103}},\ \bibinfo
  {pages} {033325} (\bibinfo {year} {2021})}\BibitemShut {NoStop}%
\bibitem [{\citenamefont {Tzortzakakis}\ \emph {et~al.}(2021)\citenamefont
  {Tzortzakakis}, \citenamefont {Makris}, \citenamefont {Szameit},\ and\
  \citenamefont {Economou}}]{tzortzakakis2021transport}%
  \BibitemOpen
  \bibfield  {author} {\bibinfo {author} {\bibfnamefont {A.~F.}\ \bibnamefont
  {Tzortzakakis}}, \bibinfo {author} {\bibfnamefont {K.~G.}\ \bibnamefont
  {Makris}}, \bibinfo {author} {\bibfnamefont {A.}~\bibnamefont {Szameit}},\
  and\ \bibinfo {author} {\bibfnamefont {E.~N.}\ \bibnamefont {Economou}},\
  }\bibfield  {title} {\bibinfo {title} {Transport and spectral features in
  non-hermitian open systems},\ }\href
  {https://doi.org/10.1103/PhysRevResearch.3.013208} {\bibfield  {journal}
  {\bibinfo  {journal} {Phys. Rev. Res.}\ }\textbf {\bibinfo {volume} {3}},\
  \bibinfo {pages} {013208} (\bibinfo {year} {2021})}\BibitemShut {NoStop}%
\bibitem [{\citenamefont {Zeng}\ \emph {et~al.}(2017)\citenamefont {Zeng},
  \citenamefont {Chen},\ and\ \citenamefont {L\"u}}]{zeng2017anderson}%
  \BibitemOpen
  \bibfield  {author} {\bibinfo {author} {\bibfnamefont {Q.-B.}\ \bibnamefont
  {Zeng}}, \bibinfo {author} {\bibfnamefont {S.}~\bibnamefont {Chen}},\ and\
  \bibinfo {author} {\bibfnamefont {R.}~\bibnamefont {L\"u}},\ }\bibfield
  {title} {\bibinfo {title} {Anderson localization in the non-hermitian
  aubry-andr\'e-harper model with physical gain and loss},\ }\href
  {https://doi.org/10.1103/PhysRevA.95.062118} {\bibfield  {journal} {\bibinfo
  {journal} {Phys. Rev. A}\ }\textbf {\bibinfo {volume} {95}},\ \bibinfo
  {pages} {062118} (\bibinfo {year} {2017})}\BibitemShut {NoStop}%
\bibitem [{\citenamefont {Zhai}\ \emph {et~al.}(2020)\citenamefont {Zhai},
  \citenamefont {Yin},\ and\ \citenamefont {Huang}}]{zhai2020many}%
  \BibitemOpen
  \bibfield  {author} {\bibinfo {author} {\bibfnamefont {L.-J.}\ \bibnamefont
  {Zhai}}, \bibinfo {author} {\bibfnamefont {S.}~\bibnamefont {Yin}},\ and\
  \bibinfo {author} {\bibfnamefont {G.-Y.}\ \bibnamefont {Huang}},\ }\bibfield
  {title} {\bibinfo {title} {Many-body localization in a non-hermitian
  quasiperiodic system},\ }\href {https://doi.org/10.1103/PhysRevB.102.064206}
  {\bibfield  {journal} {\bibinfo  {journal} {Phys. Rev. B}\ }\textbf {\bibinfo
  {volume} {102}},\ \bibinfo {pages} {064206} (\bibinfo {year}
  {2020})}\BibitemShut {NoStop}%
\bibitem [{\citenamefont {Lee}(2016)}]{Lee}%
  \BibitemOpen
  \bibfield  {author} {\bibinfo {author} {\bibfnamefont {T.~E.}\ \bibnamefont
  {Lee}},\ }\bibfield  {title} {\bibinfo {title} {Anomalous edge state in a
  non-hermitian lattice},\ }\href
  {https://doi.org/10.1103/PhysRevLett.116.133903} {\bibfield  {journal}
  {\bibinfo  {journal} {Phys. Rev. Lett.}\ }\textbf {\bibinfo {volume} {116}},\
  \bibinfo {pages} {133903} (\bibinfo {year} {2016})}\BibitemShut {NoStop}%
\bibitem [{\citenamefont {Martinez~Alvarez}\ \emph {et~al.}(2018)\citenamefont
  {Martinez~Alvarez}, \citenamefont {Barrios~Vargas},\ and\ \citenamefont
  {Foa~Torres}}]{Torres}%
  \BibitemOpen
  \bibfield  {author} {\bibinfo {author} {\bibfnamefont {V.~M.}\ \bibnamefont
  {Martinez~Alvarez}}, \bibinfo {author} {\bibfnamefont {J.~E.}\ \bibnamefont
  {Barrios~Vargas}},\ and\ \bibinfo {author} {\bibfnamefont {L.~E.~F.}\
  \bibnamefont {Foa~Torres}},\ }\bibfield  {title} {\bibinfo {title}
  {Non-hermitian robust edge states in one dimension: Anomalous localization
  and eigenspace condensation at exceptional points},\ }\href
  {https://doi.org/10.1103/PhysRevB.97.121401} {\bibfield  {journal} {\bibinfo
  {journal} {Phys. Rev. B}\ }\textbf {\bibinfo {volume} {97}},\ \bibinfo
  {pages} {121401} (\bibinfo {year} {2018})}\BibitemShut {NoStop}%
\bibitem [{\citenamefont {Lee}\ \emph {et~al.}(2019)\citenamefont {Lee},
  \citenamefont {Li},\ and\ \citenamefont {Gong}}]{Lee_2019}%
  \BibitemOpen
  \bibfield  {author} {\bibinfo {author} {\bibfnamefont {C.~H.}\ \bibnamefont
  {Lee}}, \bibinfo {author} {\bibfnamefont {L.}~\bibnamefont {Li}},\ and\
  \bibinfo {author} {\bibfnamefont {J.}~\bibnamefont {Gong}},\ }\bibfield
  {title} {\bibinfo {title} {Hybrid higher-order skin-topological modes in
  nonreciprocal systems},\ }\href
  {https://doi.org/10.1103/PhysRevLett.123.016805} {\bibfield  {journal}
  {\bibinfo  {journal} {Phys. Rev. Lett.}\ }\textbf {\bibinfo {volume} {123}},\
  \bibinfo {pages} {016805} (\bibinfo {year} {2019})}\BibitemShut {NoStop}%
\bibitem [{\citenamefont {Mu}\ \emph {et~al.}(2022)\citenamefont {Mu},
  \citenamefont {Zhou}, \citenamefont {Li},\ and\ \citenamefont
  {Gong}}]{PhysRevB.105.205402}%
  \BibitemOpen
  \bibfield  {author} {\bibinfo {author} {\bibfnamefont {S.}~\bibnamefont
  {Mu}}, \bibinfo {author} {\bibfnamefont {L.}~\bibnamefont {Zhou}}, \bibinfo
  {author} {\bibfnamefont {L.}~\bibnamefont {Li}},\ and\ \bibinfo {author}
  {\bibfnamefont {J.}~\bibnamefont {Gong}},\ }\bibfield  {title} {\bibinfo
  {title} {Non-hermitian pseudo mobility edge in a coupled chain system},\
  }\href {https://doi.org/10.1103/PhysRevB.105.205402} {\bibfield  {journal}
  {\bibinfo  {journal} {Phys. Rev. B}\ }\textbf {\bibinfo {volume} {105}},\
  \bibinfo {pages} {205402} (\bibinfo {year} {2022})}\BibitemShut {NoStop}%
\bibitem [{\citenamefont {Rossignolo}\ and\ \citenamefont
  {Dell'Anna}(2019{\natexlab{a}})}]{PhysRevB.99.054211}%
  \BibitemOpen
  \bibfield  {author} {\bibinfo {author} {\bibfnamefont {M.}~\bibnamefont
  {Rossignolo}}\ and\ \bibinfo {author} {\bibfnamefont {L.}~\bibnamefont
  {Dell'Anna}},\ }\bibfield  {title} {\bibinfo {title} {Localization
  transitions and mobility edges in coupled aubry-andr\'e chains},\ }\href
  {https://doi.org/10.1103/PhysRevB.99.054211} {\bibfield  {journal} {\bibinfo
  {journal} {Phys. Rev. B}\ }\textbf {\bibinfo {volume} {99}},\ \bibinfo
  {pages} {054211} (\bibinfo {year} {2019}{\natexlab{a}})}\BibitemShut
  {NoStop}%
\bibitem [{\citenamefont {Roy}\ \emph {et~al.}(2022)\citenamefont {Roy},
  \citenamefont {Chattopadhyay}, \citenamefont {Mishra},\ and\ \citenamefont
  {Basu}}]{PhysRevB.105.214203}%
  \BibitemOpen
  \bibfield  {author} {\bibinfo {author} {\bibfnamefont {S.}~\bibnamefont
  {Roy}}, \bibinfo {author} {\bibfnamefont {S.}~\bibnamefont {Chattopadhyay}},
  \bibinfo {author} {\bibfnamefont {T.}~\bibnamefont {Mishra}},\ and\ \bibinfo
  {author} {\bibfnamefont {S.}~\bibnamefont {Basu}},\ }\bibfield  {title}
  {\bibinfo {title} {Critical analysis of the reentrant localization transition
  in a one-dimensional dimerized quasiperiodic lattice},\ }\href
  {https://doi.org/10.1103/PhysRevB.105.214203} {\bibfield  {journal} {\bibinfo
   {journal} {Phys. Rev. B}\ }\textbf {\bibinfo {volume} {105}},\ \bibinfo
  {pages} {214203} (\bibinfo {year} {2022})}\BibitemShut {NoStop}%
\bibitem [{\citenamefont {Rossignolo}\ and\ \citenamefont
  {Dell'Anna}(2019{\natexlab{b}})}]{Rossignolo}%
  \BibitemOpen
  \bibfield  {author} {\bibinfo {author} {\bibfnamefont {M.}~\bibnamefont
  {Rossignolo}}\ and\ \bibinfo {author} {\bibfnamefont {L.}~\bibnamefont
  {Dell'Anna}},\ }\bibfield  {title} {\bibinfo {title} {Localization
  transitions and mobility edges in coupled aubry-andr\'e chains},\ }\href
  {https://doi.org/10.1103/PhysRevB.99.054211} {\bibfield  {journal} {\bibinfo
  {journal} {Phys. Rev. B}\ }\textbf {\bibinfo {volume} {99}},\ \bibinfo
  {pages} {054211} (\bibinfo {year} {2019}{\natexlab{b}})}\BibitemShut
  {NoStop}%
\bibitem [{\citenamefont {Jiang}\ \emph {et~al.}(2019)\citenamefont {Jiang},
  \citenamefont {Lang}, \citenamefont {Yang}, \citenamefont {Zhu},\ and\
  \citenamefont {Chen}}]{Jiang}%
  \BibitemOpen
  \bibfield  {author} {\bibinfo {author} {\bibfnamefont {H.}~\bibnamefont
  {Jiang}}, \bibinfo {author} {\bibfnamefont {L.-J.}\ \bibnamefont {Lang}},
  \bibinfo {author} {\bibfnamefont {C.}~\bibnamefont {Yang}}, \bibinfo {author}
  {\bibfnamefont {S.-L.}\ \bibnamefont {Zhu}},\ and\ \bibinfo {author}
  {\bibfnamefont {S.}~\bibnamefont {Chen}},\ }\bibfield  {title} {\bibinfo
  {title} {Interplay of non-hermitian skin effects and anderson localization in
  nonreciprocal quasiperiodic lattices},\ }\href
  {https://doi.org/10.1103/PhysRevB.100.054301} {\bibfield  {journal} {\bibinfo
   {journal} {Phys. Rev. B}\ }\textbf {\bibinfo {volume} {100}},\ \bibinfo
  {pages} {054301} (\bibinfo {year} {2019})}\BibitemShut {NoStop}%
\bibitem [{\citenamefont {Kokkinakis}\ \emph {et~al.}(2024)\citenamefont
  {Kokkinakis}, \citenamefont {Makris},\ and\ \citenamefont
  {Economou}}]{kokkinakis2024andersonlocalizationversushopping}%
  \BibitemOpen
  \bibfield  {author} {\bibinfo {author} {\bibfnamefont {E.~T.}\ \bibnamefont
  {Kokkinakis}}, \bibinfo {author} {\bibfnamefont {K.~G.}\ \bibnamefont
  {Makris}},\ and\ \bibinfo {author} {\bibfnamefont {E.~N.}\ \bibnamefont
  {Economou}},\ }\bibfield  {title} {\bibinfo {title} {Anderson localization
  versus hopping asymmetry in a disordered lattice},\ }\bibfield  {journal}
  {\bibinfo  {journal} {arXiv}\ }\href
  {https://doi.org/https://doi.org/10.48550/arXiv.2407.10746}
  {https://doi.org/10.48550/arXiv.2407.10746} (\bibinfo {year}
  {2024})\BibitemShut {NoStop}%
\bibitem [{\citenamefont {Kraus}\ \emph {et~al.}(2012)\citenamefont {Kraus},
  \citenamefont {Lahini}, \citenamefont {Ringel}, \citenamefont {Verbin},\ and\
  \citenamefont {Zilberberg}}]{PhysRevLett.109.106402}%
  \BibitemOpen
  \bibfield  {author} {\bibinfo {author} {\bibfnamefont {Y.~E.}\ \bibnamefont
  {Kraus}}, \bibinfo {author} {\bibfnamefont {Y.}~\bibnamefont {Lahini}},
  \bibinfo {author} {\bibfnamefont {Z.}~\bibnamefont {Ringel}}, \bibinfo
  {author} {\bibfnamefont {M.}~\bibnamefont {Verbin}},\ and\ \bibinfo {author}
  {\bibfnamefont {O.}~\bibnamefont {Zilberberg}},\ }\bibfield  {title}
  {\bibinfo {title} {Topological states and adiabatic pumping in
  quasicrystals},\ }\href {https://doi.org/10.1103/PhysRevLett.109.106402}
  {\bibfield  {journal} {\bibinfo  {journal} {Phys. Rev. Lett.}\ }\textbf
  {\bibinfo {volume} {109}},\ \bibinfo {pages} {106402} (\bibinfo {year}
  {2012})}\BibitemShut {NoStop}%
\end{thebibliography}%
\end{document}